\documentclass[aps,prx,10pt,twocolumn,superscriptaddress,noeprint]{revtex4-2}
\usepackage[utf8]{inputenc}
\usepackage[unicode,pdfusetitle,colorlinks,urlcolor=blue,linkcolor=blue,citecolor=blue]{hyperref}
\hypersetup{pdfauthor={J. H. Béjanin}}
\usepackage{graphicx}
\usepackage{physics,siunitx,amsmath}
\usepackage[ruled,lined]{algorithm2e}

\newcommand{\T}[1]{T\textsubscript{#1}}

\begin{document}

\title{Resonant Coupling Parameter Estimation with Superconducting Qubits}

\newcommand{\IQC}{%
  \affiliation{%
    Institute for Quantum Computing,
    University of Waterloo,
    200 University Avenue West,
    Waterloo, Ontario N2L 3G1, Canada
  }
}

\newcommand{\WatPhys}{%
  \affiliation{%
    Department of Physics and Astronomy,
    University of Waterloo,
    200 University Avenue West,
    Waterloo, Ontario N2L 3G1, Canada
  }
}

\newcommand{\MQ}{%
  \affiliation{%
    Department of Physics and Astronomy,
    Macquarie University,
    Sydney, NSW 2109, Australia
  }
}

\newcommand{\ARC}{%
  \affiliation{%
  ARC Centre of Excellence in Engineered Quantum Systems,
    Macquarie University,
    Sydney, NSW 2109, Australia
  }
}

\author{J.~H.~Béjanin}
\IQC \WatPhys
\author{C.~T.~Earnest} \IQC \WatPhys
\author{Y.~R.~Sanders} \MQ \ARC
\author{M.~Mariantoni}
\email[Corresponding author:~]{matteo.mariantoni@uwaterloo.ca}
\IQC \WatPhys

\date{\today}

\begin{abstract}
Today's quantum computers are comprised of tens of qubits interacting with each other and the environment in increasingly complex networks. In order to achieve the best possible performance when operating such systems, it is necessary to have accurate knowledge of all parameters in the quantum computer Hamiltonian. In this article, we demonstrate theoretically and experimentally a method to efficiently learn the parameters of resonant interactions for quantum computers consisting of frequency-tunable superconducting qubits. Such interactions include, for example, those to other qubits, resonators, two-level state defects, or other unwanted modes. Our method is based on a significantly improved swap spectroscopy calibration and consists of an \emph{offline} data collection algorithm, followed by an \emph{online} Bayesian learning algorithm. The purpose of the offline algorithm is to detect and roughly estimate resonant interactions from a state of zero knowledge. It produces a square-root reduction in the number of measurements. The online algorithm subsequently refines the estimate of the parameters to comparable accuracy as traditional swap spectroscopy calibration, but in constant time. We perform an experiment implementing our technique with a superconducting qubit. By combining both algorithms, we observe a reduction of the calibration time by one order of magnitude. We believe the method investigated will improve present medium-scale superconducting quantum computers and will also scale up to larger systems. Finally, the two algorithms presented here can be readily adopted by communities working on different physical implementations of quantum computing architectures.
\end{abstract}

\maketitle

\section{Introduction}

Quantum computing architectures based on different types of physical qubits have been investigated since the late 1990s~\cite{Ladd2010}. Although quantum computers will be lacking fault tolerance for the near future~\cite{Gottesman2010,Fowler2012}, a variety of potential applications have been devised for such systems. Examples include solving optimization problems\cite{Kechedzhi2018,Somma2008}, machine learning~\cite{Farhi2018,McClean2018,Cong2019,Bravyi2018}, materials science and chemistry~\cite{Aspuru-Guzik2005,Peruzzo2014,Hempel2018}, and even commercializable technologies~\cite{Mohseni2017}.

Small- and medium-scale quantum computers built out of superconducting circuits~\cite{Clarke2008,Wendin2017,Kockum2019,Mariantoni2020} and comprising up to a few tens of physical qubits have been in operation for the past few years~\cite{Mariantoni2011,Lucero2012,Corcoles2015,Riste2015,Barends2014,Kelly2015}. The two main reasons for the success of superconducting technologies in the quantum computing arena are the pre-existing facilities for scaling up fabrication, due to methods being similar to silicon~(Si) technology, as well as the favorable coherence-to-gate-time ratio. For state-of-the-art superconducting qubits with very short gate times of less than~\SI{100}{\nano\second} (e.g., see Ref.~\onlinecite{Barends2013}), such a ratio reaches values close to~$1000$. This means that thousands of one- and two-qubit gates can be performed within the qubit lifetime, with fidelities in excess of~\SI{99.4}{\percent}~\cite{Barends2014}. Very recently, a major milestone in quantum computing has been reached~—~quantum supremacy~—~where a $53$-qubit superconducting quantum computer has been utilized to sample the output of a pseudo-random quantum circuit~\cite{Arute2019}. As a result of these advances, quantum computing has transitioned into the so-called noisy intermediate scale quantum~(NISQ) era~\cite{Preskill2018}.

Despite these successes, the high-fidelity operation of medium- and large-scale quantum computers is accompanied by the daunting task of calibrating numerous physical qubits. In particular, calibrating tunable qubits requires the estimation of resonant interaction parameters such as the resonance frequency and coupling coefficient between pairs of interacting qubits~\cite{DiCarlo2009,Yamamoto2010}, a qubit and a resonator~\cite{Mariantoni2011}, two-level state~(TLS) defects~\cite{Muller2019,Earnest2018,Moeed2019}, or substrate and box modes~\cite{McConkey2018}. These calibrations are necessary to implement two-qubit gates and avoid loss of quantum information due to spurious interactions leading to coherent or incoherent errors. TLS defects, especially, are a pervasive source of errors in superconducting architectures which must be remediated~\cite{Klimov2018}.

In this article, we study theoretically and demonstrate experimentally a data-efficient and automated method for identifying and estimating the parameters of resonant interactions based on \emph{swap spectroscopy}~\cite{Mariantoni2011a,Mariantoni2011}. We realize swap spectroscopy by performing energy relaxation time~\T1 measurements of a frequency-tunable Xmon transmon qubit~\cite{Barends2013} at different qubit frequencies. The identification and estimation method is divided into two parts: an \emph{offline} data collection algorithm~\cite{Sanders2016} and an \emph{online} Bayesian learning algorithm~\cite{Stenberg2014,Granade2012}. Both algorithms are based on the dynamics of interacting quantum systems. The former is used from a state of zero knowledge to roughly identify resonance parameters; the latter focuses on improving the estimate of those parameters. In this context, the term ``online'' means that measurements taken during the execution of the algorithm inform the subsequent ones. For the ``offline'' method, the execution of the entire algorithm is predetermined.

By means of our parameter-estimation method, we can to shorten the calibration time of an Xmon transmon qubit significantly. The \emph{offline} data collection algorithm makes it possible to reduce the number of measurements by a square-root factor when compared to a traditional swap spectroscopy calibration. In our experiment, this algorithm takes~$\approx\SI{30}{\minute}$ to detect resonances in a $\SI{1}{\giga\hertz}$ bandwidth; one order of magnitude less time that it would take previously. The online Bayesian learning algorithm runs in $\approx\SI{25}{\second}$ per resonance and improves the estimation accuracy to be equivalent to that of a high-resolution traditional swap spectroscopy.

Our method is not confined to the realm of superconducting quantum computing. In fact, it can easily be adopted by practitioners working on different physical implementations of quantum computing architectures such as trapped ions and semiconductor qubits~\cite{Ladd2010}.

This article is organized as follows: in Sec.~\ref{sec:calibration}, we explain qubit calibration in frequency-tunable architectures; in Sec.~\ref{sec:naiveswapspec} we summarize the working principle of traditional swap spectroscopy, explaining why it is inefficient; in Sec.~\ref{sec:octavesampling}, we introduce the offline octave sampling algorithm (Subsec.~\ref{subs:octavetheory}) and show our experimental implementation of the method to find the interaction parameters between an Xmon transmon qubit and three resonance modes~(RMs) (Subsec.~\ref{subs:octaveresutls}); in Sec.~\ref{sec:bayesianlearning}, we explain the online Bayesian learning algorithm (Subsec.~\ref{subs:bayesiantheory}) and demonstrate its performance for refining the estimate of the resonance parameters (Subsec.~\ref{subs:bayesianresults}); in Sec.~\ref{sec:discussion} we discuss additional concerns with the algorithms and the relevance of our methods for quantum computing; finally, in Sec.~\ref{sec:conclusion}, we provide an outlook and conclusions.

\section{Qubit Calibration in Frequency-Tunable Architectures}
\label{sec:calibration}

A fundamental requirement to the operation of a quantum computer is the proper calibration of the physical qubits in the system. This calibration includes many specific operations. One of the most basic tasks, for example, is to run a Rabi experiment on each qubit. This allows the determination of some experimental parameters needed to set up, e.g., a~$\pi$-pulse and perform a measurement. Once this first task has been realized, further experiments can refine the knowledge of the pulse amplitude, rotation axis, measurement parameters, etc. Finally, a full calibration requires knowing the precise parameters of the total system Hamiltonian, allowing for the systematic optimization of the fidelity of one- and two-qubit gates as well as measurement.

In a frequency-tunable superconducting qubit architecture such as the Google architecture~\cite{Arute2019} or the one used in this work, an additional degree of freedom must be considered during calibration: the qubit frequency. Xmon transmon qubits are one example of tunable qubits~\cite{Barends2013}. In this design, an on-chip capacitive island made from aluminum~(Al) is coupled in parallel to a DC~superconducting quantum interference device~(SQUID) comprised of two Josephson tunnel junctions in parallel, forming a superconducting loop~\cite{Mariantoni2020}. Simply put, an Xmon transmon qubit is a quantum \emph{anharmonic} oscillator, characterized by a non-equally spaced ladder of quantum states. The frequency (i.e., energy) difference~$f_q$ between the energy ground state~$\ket{\text{g}}$ and first excited state~$\ket{\text{e}}$ differs from that between~$\ket{\text{e}}$ and the second excited state~$\ket{\text{f}}$ by the so-called qubit anharmonicity~$\alpha$~\cite{Koch2007}. The qubit transition frequency~$f_q$ is controlled \textit{in situ} by applying a local external magnetic flux that threads the~DC~SQUID, tuning the Josephson energy~$E_{\text{J}}$ and therefore the level separation. This degree of freedom leads to a few distinct advantages to the operation of a quantum computer.

For instance, frequency tunability allows for tunable qubit-qubit interactions because the effective coupling strength between two qubits depends on the frequency difference between them. This enables the implementation of several types of two-qubit gates such as the controlled-phase~(CPHASE) gate, which takes advantage of state~$\ket{\text{f}}$ as an auxiliary state~\cite{Haack2010,DiCarlo2009,Yamamoto2010,Mariantoni2011}, as well as the~$\sqrt{i\text{SWAP}}$ and $i\text{SWAP}$ gates~\cite{Mariantoni2011}. In addition, setting the frequency of spatially neighboring qubits away from each other helps avoid control crosstalk and frequency crowding issues, with the latter being endemic in fixed-frequency systems~\cite{Krantz2019}.

Another advantage inherent to frequency tunable architectures has to do with energy relaxation. On-chip superconducting qubits interact with a distribution of~TLS defects, which are present in the various amorphous dielectric materials surrounding the qubit metallic structures (e.g., Si and Al oxides). While the microscopic origin of TLS defects is still under debate, their effect on the qubit can be broadly categorized either as causing frequency noise or excitation noise~\cite{Muller2019}. In particular, TLS defects interacting semi-resonantly with a qubit can coherently or incoherently exchange energy with it resulting in a lossy channel. The ability to set the frequency of a qubit away from that of TLS defects is therefore desirable, and realizable only with tunability.

Calibrating qubits to implement two-qubit gates or to avoid~TLS defects is a~\emph{parameter estimation} problem. We need to determine the Hamiltonian parameters that define the resonant interactions between a qubit and another system. In all the aforementioned cases, two parameters must be found: the resonance frequency and coupling strength of the interaction.

Historically, this kind of calibration has been realized using swap spectroscopy. Unfortunately, traditional swap spectroscopy is inefficient in the amount of data it requires, and therefore slow. This is inconvenient for multiple reasons. First, as the number of qubits in a system grows, so does the number of calibrations that must be performed. This is particularly relevant to qubit-qubit coupling calibration, which cannot be performed in parallel on all qubits. Second, TLS defects in the environment are known to fluctuate over time~\cite{Klimov2018,Schlor2019,Burnett2019,Moeed2019}. Similarly, $f_q$ itself can shift in time. The identification of resonant interactions must therefore be repeated at regular intervals. We thus require a robust, accurate, and time-efficient method to identify the parameters associated with resonant interactions.

In order to test our parameter-estimation method and compare it against traditional swap spectroscopy, we use an Xmon transmon qubit and make it interact with a synthesized resonance mode. The mode, which emulates the interaction with another qubit, resonator, or TLS defect, is created by applying a coherent drive with a microwave source to the qubit under test. The synthesized resonance mode is a convenient and flexible tool to test our method since we can arbitrarily change its \emph{resonance frequency} by tuning the source frequency as well as its \emph{coupling strength} by changing the emitted source power.

\section{Traditional Swap Spectroscopy}
\label{sec:naiveswapspec}

Swap spectroscopy is an experimental method that allows exploring the environment of a qubit at various frequencies by using the qubit itself as a probe. Traditionally, swap spectroscopy has been used to select the operating frequency of qubits, making it possible to avoid TLS defects or regions of low~\T1. Additionally, it has been used to explore resonant interactions, such as those with other qubits~\cite{Yamamoto2010} or resonators~\cite{Mariantoni2011a}. Performing swap spectroscopy requires a minimally calibrated qubit and, thus, is suitable as a tune-up experiment.

\begin{figure}[b]
\includegraphics[width=8cm,height=2cm]{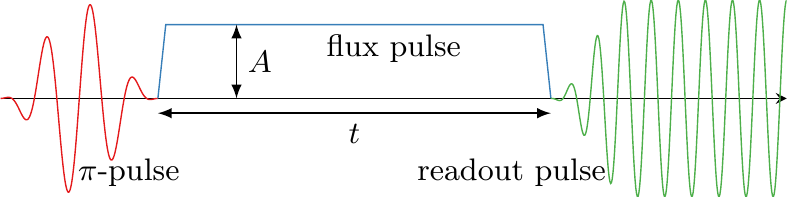}
\caption{Pulse sequence for a swap spectroscopy experiment. The initial $\pi$-pulse (red) excites the qubit, which is initialized at the so-called idle frequency. The flux pulse (blue) changes the qubit transition frequency~$f_q$ to the probe frequency~$f_\text{p}(A)$ for a duration~$t$. The qubit is then measured (green) after being set back to its idle frequency.}
\label{fig:pulsesequence}
\end{figure}

In a swap spectroscopy experiment the qubit is initialized at the so-called \emph{idle frequency}. A~$\pi$-pulse is then applied to the qubit, energizing it from~$\ket{\text{g}}$ to $\ket{\text{e}}$. At the end of the~$\pi$-pulse, a flux pulse is applied to the~DC~SQUID in order to tune the qubit to a different frequency, the \emph{probe frequency}~$f_\text{p}(A)$. This procedure requires knowledge of the correspondence between the qubit frequency and pulse amplitude~$A$, which can be calibrated via regular pulse spectroscopy~(see App.~\ref{ap:fluxpulse}). After a time~$t$, the flux pulse is turned off and the qubit is measured back at the idle frequency. This pulse sequence is illustrated in Fig.~\ref{fig:pulsesequence}. Note that using a flux pulse to set $f_\text{p}(A)$ presents advantages over quasi-statically changing the idle qubit frequency by means of a DC current to the SQUID. Namely, it avoids recalibrating the~$\pi$ pulse and measurement pulse at each qubit frequency.

In a traditional experiment, $t$ and $f_\text{p}(A)$ are swept linearly over a desired range and the qubit is measured at each point, recording how many shots correspond to an excited or ground state, $n_e$ or $n_g$, respectively. As a result of measuring the qubit in the Z~basis, swap spectroscopy is mostly insensitive to qubit dephasing.

\begin{figure*}[t]
\begin{minipage}{0.5\linewidth}
    \includegraphics[width=\textwidth]{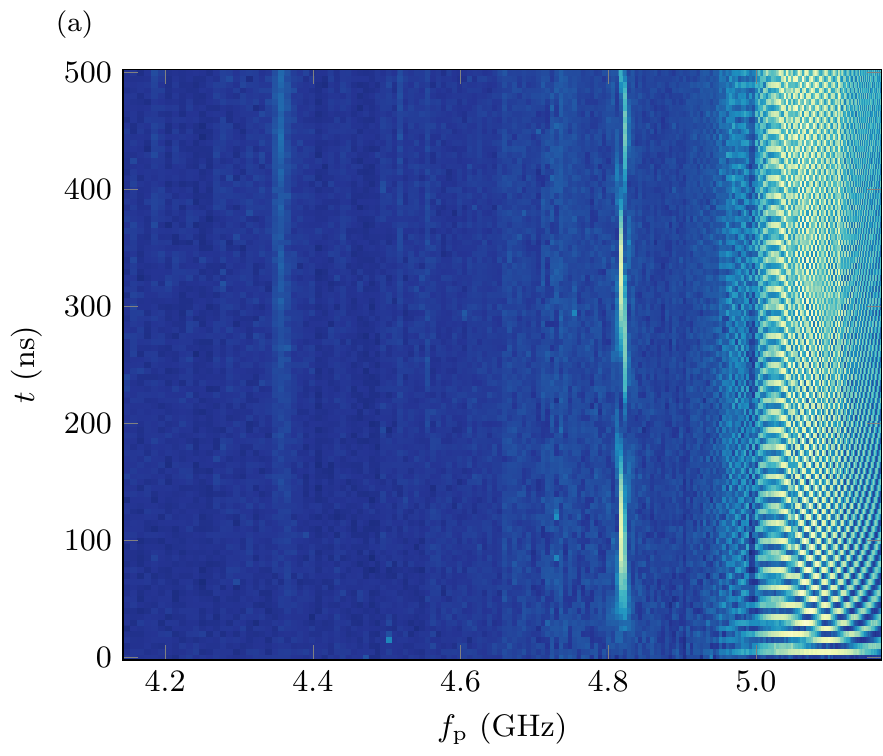}
\end{minipage}%
\begin{minipage}{0.5\linewidth}
    \includegraphics[width=\textwidth]{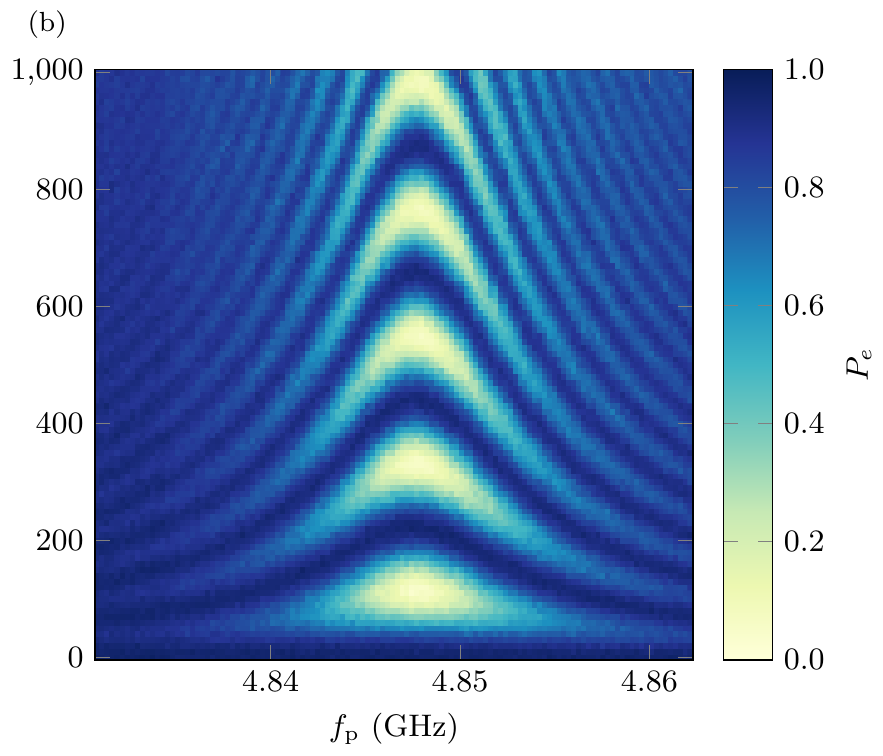}
\end{minipage}%
\caption{Traditional swap spectroscopy experiments. The $x$-axis shows the probe frequency of the qubit, which is set by the amplitude of the flux pulse applied to the DC SQUID. The $y$-axis indicates the length of the flux pulse before measurement. (a) Distinct features are visible in the full spectrum, including two chevron patterns, around~\SI{4.8}{\giga\hertz} and \SI{5.1}{\giga\hertz}. (b) Zoom in the region with the slower coherent chevron pattern due to the synthesized resonance mode. A wide frequency scan is needed to see if and where there are resonance modes, as in (a), but a more detailed experiment, as in (b), is needed to properly estimate the resonance parameters.}
\label{fig:swapspec}
\end{figure*}

Figure~\ref{fig:swapspec}~(a) shows the result of a typical swap spectroscopy experiment, with data taken between~\SI{4.146}{\giga\hertz} and~\SI{5.170}{\giga\hertz} and for times up to~\SI{500}{\nano\second}. Resonant couplings appear as oscillations, or chevron patterns, of the measured average population~$P_e = n_e/(n_e+n_g)$ in time. For example, on the far right of the spectrum it is possible to observe very fast oscillations, corresponding to a strong coupling of~$g\approx\SI{40}{\mega\hertz}$ between the qubit and the measurement resonator (see Appendix~\ref{ap:setup} for details on the sample layout and experimental setup). To the left of the resonator we observe a slower oscillation corresponding to a weaker interaction between the qubit and synthesized resonance mode. Finally, at an even lower frequency, around~\SI{4.35}{\giga\hertz}, we observe a ``streaky'' structure. In this region, the qubit excitation is lost faster than elsewhere, and we cannot observe any oscillation.

The features observed in Fig.~\ref{fig:swapspec}~(a) demonstrate a selection of possible resonant interactions: strong interactions, where~$g\gg1/T_1$ resulting in multiple coherent oscillation cycles and weak interactions appearing as regions of lower~\T1. Neither is ideal for the operation of a qubit. In the case shown in Fig.~\ref{fig:swapspec}, the best choice for the qubit idle frequency is around~\SI{4.6}{\giga\hertz}; far away from any unwanted couplings.

The data in Fig.~\ref{fig:swapspec}~(a) gives us a rough idea about the parameters of any possible resonance modes coupled to the qubit within the measured spectrum. It is hard to tell, however, that there are in fact \emph{two} resonance modes at~\SI{4.8}{\giga\hertz}, or what the frequency of the oscillation for the resonator is. A more detailed scan, such as the one in Fig.~\ref{fig:swapspec}~(b), might be necessary to estimate the parameters with sufficient accuracy. Traditional swap spectroscopy, with data taken in a linear grid, is an effective method to detect and estimate resonance modes. We will show in Sec.~\ref{sec:octavesampling}, however, that the traditional method is inefficient, and that there exists a much better way to perform this task.

\section{Offline Octave Sampling}
\label{sec:octavesampling}

The offline octave sampling algorithm has a similar objective as swap spectroscopy, i.e., to determine if there are any systems interacting resonantly with the qubit and provide an estimate for their coupling parameters. However, we want to achieve this purpose in a more efficient fashion by acquiring fewer data and therefore saving time. Note that the basic experiment employed, that is the pulse sequence illustrated in Fig.~\ref{fig:pulsesequence}, is the same as for swap spectroscopy. The difference lies in how the data is sampled. Whereas traditional swap spectroscopy samples the frequency-time space in a regular grid, octave sampling takes advantage of TLS dynamics to acquire as few data as possible.

\subsection{Theoretical Method}
\label{subs:octavetheory}

\begin{figure*}[t]
\begin{minipage}{0.4\linewidth}
  \includegraphics[width=\linewidth]{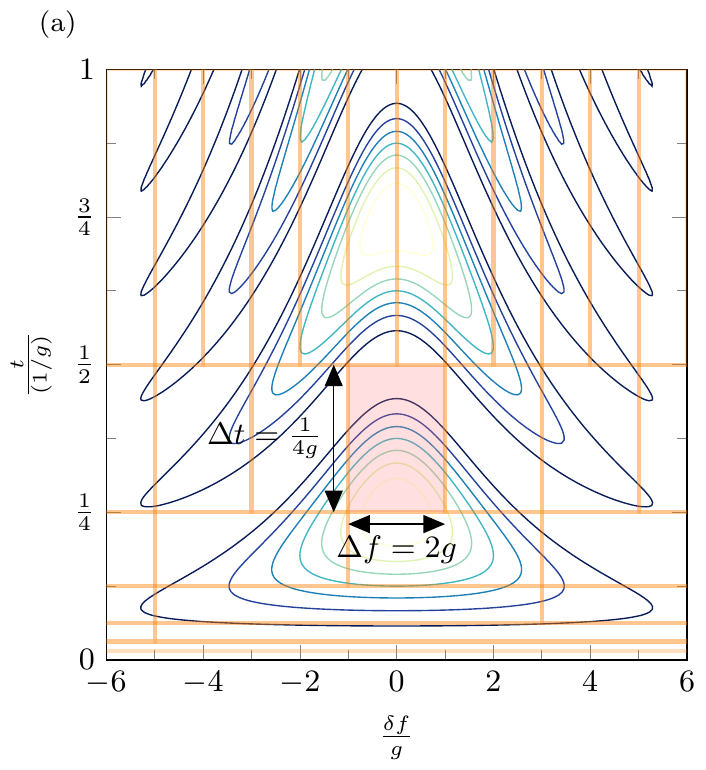}
\end{minipage}%
\begin{minipage}{0.6\linewidth}
  \includegraphics[width=\linewidth]{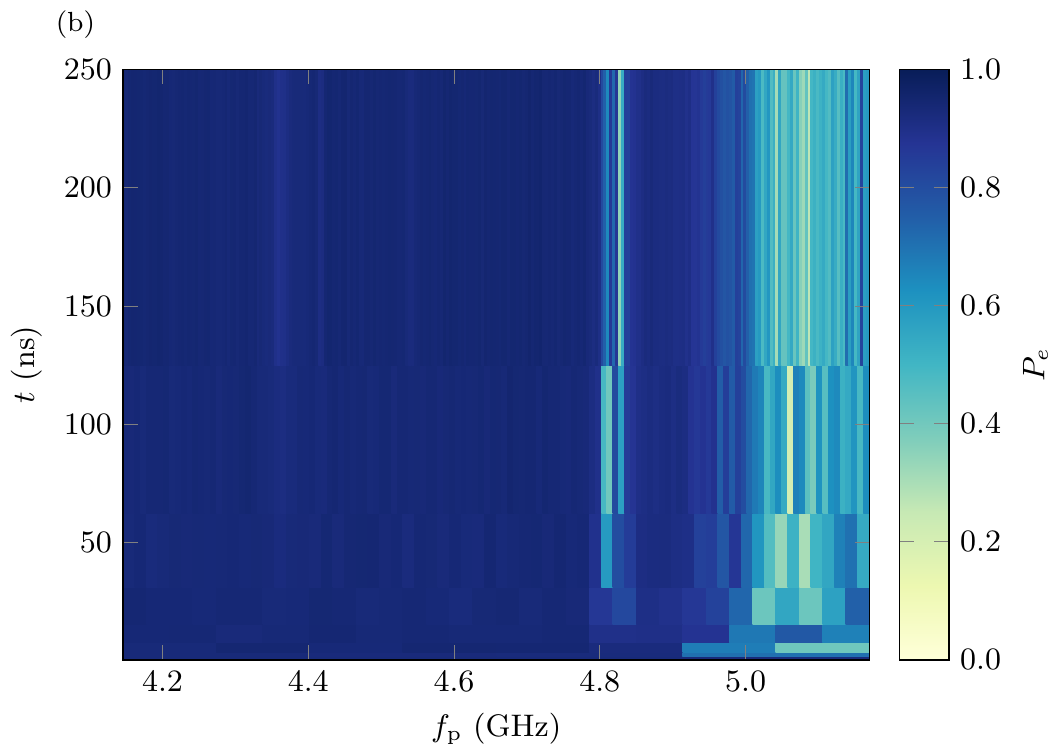}
\end{minipage}%
\caption{Offline octave sampling. (a) Contour plot of the probability of finding the qubit in~$\ket{\text{e}}$ [see Eq.~(\ref{eq:theoryPe})] as a function of time~$t$ and frequency detuning~$\Delta f$; both axes are normalized by the coupling strength~$g$. (b) Swap spectroscopy experiment with octave sampling. For each octave the bin width is halved and the bin length is doubled. The color of each bin represents the average value of the measured $P_e$ over~$n_s=5$ samples.}
\label{fig:octaves}
\end{figure*}

In order to explain the data collection strategy, we analyze the time dynamics of the systems at play. Since we are searching for resonant interactions with a qubit, we work in a single-excitation manifold ($\ket{\text{g}} \leftrightarrow \ket{\text{e}}$). Thus, even if a system is characterized by multiple energy levels (e.g., a resonator), we can still treat it as a two-level system, regardless of the underlying physics. This is our working assumption throughout the rest of the article.

Note that we can probe the environment of a qubit within a different single-excitation manifold. For example, if we have an anharmonic oscillator as a probe instead of a true qubit, we can populate the second excited state and look for systems coupled to the~$\ket{\text{e}} \leftrightarrow \ket{\text{f}}$ transition. This allows for the calibration of a CPHASE two-qubit gate~\cite{Yamamoto2010,Mariantoni2011,Barends2014}. In either case, because we consider the exchange of a single excitation, the effective Hamiltonian remains unchanged.

After a rotating wave approximation, the Hamiltonian of a qubit at the probe frequency~$f_\text{p}$ interacting with a resonance mode at a frequency $f_\text{RM}$ reads
\begin{equation}
\hat{H} = \frac{h f_\text{p}}{2} \hat{\sigma}_{z,1} + \frac{h f_\text{RM}}{2} \hat{\sigma}_{z,2} + h g \left( \hat{\sigma}^+_1 \hat{\sigma}^-_2 + \hat{\sigma}^-_1 \hat{\sigma}^+_2 \right),
\end{equation}
where $g$ is the coupling strength of the qubit-resonance mode interaction, $\hat{\sigma}_{z,1(2)}$ are Pauli matrices for the qubit~($1$) and resonance mode~($2$), and $\hat{\sigma}^+_{1(2)}$ and  $\hat{\sigma}^-_{1(2)}$ are rising and lowering operators for the qubit and resonance mode. We solve for the time evolution of the qubit when it is initialized in state~$\ket{\text{e}}$ and with the resonance mode starting in~$\ket{\text{g}}$. The theoretical probability of finding the qubit in the excited state after a time $t$ is then given by
\begin{equation}
\label{eq:theoryPe}
\tilde{P_e}(t) = 1 - \left(\frac{2g}{\Omega}\right)^2 \sin(2\pi \Omega t/2)^2,
\end{equation}
where~$\Omega^2 = \delta f^2+4g^2$, with~$\delta f = f_\text{p} - f_\text{RM}$. The probability $\tilde{P_e}$ of Eq.~(\ref{eq:theoryPe}) is plotted in Fig.~\ref{fig:octaves}~(a) as contours. Close to resonance, the excitation swaps between the qubit and the resonance mode with frequency $\Omega$ increasing at larger~$\delta f$, resulting in the familiar chevron pattern. Both the width of the pattern, which we quantify by the full-width half maximum of the amplitude, $4g$, and the frequency of the oscillation depend on $g$, the coupling strength. Crucially, the width depends linearly on $g$, while the period of the oscillation (and therefore the time location of the first minimum) depend on $1/g$.

With these observations in mind, we choose to divide the frequency-time space into \emph{bins} within which we take a constant number $n_s$ of swap spectroscopy measurements. Instead of naively sampling the spectrum in a uniform grid, we adapt the measurement based on the coupling strength $g$ we are trying to detect. $g$ determines the \emph{time} $t$ at which we measure and the \emph{bin size}. On the one hand, a resonance mode with large coupling strength~$g$ has a large width and a short period. For small $t$ then, we choose bins to be wide and short. On the other hand, a more weakly coupled resonance mode appears later in time and has a narrower frequency width and a longer period. In this case, the bins are longer and narrower [see Fig.~\ref{fig:octaves}~(a)]. Because we choose the size of these bins to vary by factors of two, we refer to this method as \emph{octave} sampling.

The goal of octave sampling is to detect a resonance mode by finding the \emph{first minimum} of an oscillation, where measurements give $P_e\sim0$ because the excitation has swapped into the resonance mode. In order to make this division systematic, we introduce the concept of a coupling octave, characterized by the coupling strength $g_o$, where $o$ is the octave number which ranges from $0$ to $o_f$~\cite{Sanders2016}. The final octave number $o_f$ is determined by the frequency or time resolution that is desired.

For each octave, the full frequency spectrum to be analyzed ranges between a minimum and maximum frequency $f_\text{min}$ and $f_\text{max}$. This range is divided into $2^o$ bins of equal size, with frequency width $\Delta f = 2g_o$ and time length $\Delta t = 1/(4g_o)$. The location in time of the bins' lower edge is $t = 1/(4g_o)$. One such bin, with $g_o=g$, is highlighted in red in Fig.~\ref{fig:octaves}~(a). Note that the highlighted bin is not centered on the oscillation minimum. This is because a low-$P_e$ measurement in that area corresponds to a \emph{range} of possible coupling strengths, namely, those for which $g_o/2 \leq g \leq g_o$. The resonance mode plotted in Fig.~\ref{fig:octaves}~(a) is at the upper end of this range, and, hence, at the lower edge of the bin.

The execution of the algorithm is determined by~$\Delta = f_\text{max} - f_\text{min}$. The zeroth octave consists of a single bin which spans the whole spectrum. This width corresponds to a coupling octave $g_0 = \Delta/2$, and therefore to a time, $t = 1/4g_0$. For the next octave, we divide the width of the bins by two such that the scan has twice as many bins as for the previous step. The length of the bins in time is correspondingly doubled. An example of this division is shown by the orange grid in Fig.~\ref{fig:octaves}~(a).

Explicitly, starting from the zeroth octave~$o=0$:
\begin{enumerate}
    \item Divide the frequency range in $2^o$ bins, each with width~$2g_o = \Delta/2^o$;
    \item Each bin's time length is~$t \in [ 1/4g_o , 1/2g_o ]$;
    \item Take~$n_s$ swap spectroscopy samples within each of the~$2^o$ bins, sampling uniformly in frequency and inverse uniformly in time, $k\in [1 \ldots 2^o]$,
    \begin{align*}
        f_\text{p} &\sim \mathcal{U}\left(f_\text{min}+2(k-1)g_o,f_\text{min}+2k g_o\right), \\
        t &\sim \mathcal{U}\left(2g_0,4g_o\right)^{-1},
    \end{align*}
    where the notation $X \sim \mathcal{U}(a,b)$ signifies that $X$ is drawn randomly from a continuous uniform distribution $\mathcal{U}$ between $a$ and $b$;
    \item Increment~$o$ and start over for the next octave.
\end{enumerate}

The total number of bins to be measured depends both on the size of the spectrum $\Delta$ and the final octave number $o_f$. To set $o_f$, we can choose either a maximum time $t_{o_f}=1/2g_{o_f}$, or a final frequency resolution $\Delta f_{o_f} = 2g_{o_f}$. Either one of these quantities determine the number of bins for the final octave through the octave coupling $g_{o_f}$. From the equation above, $o_f = \log_2(\Delta/\Delta f_{o_f})$; accordingly, the total number of bins is $2^{o_f+1}-1 = 2\Delta/\Delta f_{o_f}-1$. The total number of points measured is the number of bins multiplied by the number of samples per bin $n_s$.

For the same resolution, traditional swap spectroscopy divides the frequency axis in $\Delta/\Delta f_{o_f}$ points, and the time axis in $t_{o_f}/\Delta t_\text{min}$ points, where $\Delta t_\text{min}$ is the time resolution. While octave sampling reaches a time resolution of $1/\Delta$, it would be unfair to compare the traditional method directly to that number. Instead, we assume that $1/\Delta t_\text{min}$ is on the order of 100s of MHz, in order to be able to detect strong couplings, such as those to other qubits or resonators. The total number of points is then $\Delta/\Delta f_{o_f} \times t_{o_f}/\Delta t_\text{min} = 1/\Delta t_\text{min} \times \Delta/\Delta f_{o_f}^2$. We can thus see that the number of points with the traditional method scales as $\order{1/\Delta f_{o_f}^2}$, while for octave sampling it scales as $\order{1/\Delta f_{o_f}}$, a square-root improvement.

\subsection{Experimental Results}
\label{subs:octaveresutls}

The result of an experimental implementation of the octave sampling algorithm is shown in Fig.~\ref{fig:octaves}~(b), which is the efficient version of Fig.~\ref{fig:swapspec}. The frequency ranges from $f_\text{min} = \SI{4.146}{\giga\hertz}$ to $f_\text{max} = \SI{5.170}{\giga\hertz}$. We choose a final octave number $o_f=8$. Each bin is color coded according to the average excitation probability~$\overline{P_e}$ measured over~$n_s=5$ samples. We are able to discern the same features as in Fig.~\ref{fig:swapspec}, while acquiring much fewer data. For an equivalent frequency and time resolution, octave sampling requires an order of magnitude less points than traditional swap spectroscopy, making it much more efficient. For a $\SI{1024}{\mega\hertz}$ spectrum divided in $\SI{4}{\mega\hertz}$, and the ability to detect oscillations at up to $\SI{200}{\mega\hertz}$ in a $\SI{250}{\nano\second}$ time interval, traditional swap spectroscopy requires $1024/4 \times 250/2.5 = 25600$ points. The equivalent octave sampling uses 511 bins; if we take 5 samples per bin, as in Fig.~\ref{fig:octaves}~(b), we need a total of 2555 measurements.

Given the octave sampling results of Fig.~\ref{fig:octaves}~(b), we now intend to determine if there is one or more resonance modes interacting with the qubit. If there are no resonance modes at all, the qubit does not undergo any swap, and we should always measure it to be in~$\ket{\text{e}}$ with~$\overline{P_e}=1$, notwithstanding~\T1 decay. Hence, a measurement of~$P_e<1$ indicates energy loss due to a resonance mode interacting with the qubit.

In practice, however, other spurious experimental effects can lower the measured~$P_e$ below the theoretical value of one, even in the absence of a resonance mode. Those include, for instance, energy relaxation, qubit state preparation, measurement visibility, bin averaging, and statistical fluctuations. We must therefore set a threshold probability $P_t < 1$ to account for these effects.

Qubit energy relaxation unavoidably makes the qubit decay to~$\ket{\text{g}}$. In order to avoid classifying this effect as a resonance mode, the threshold should be chosen to be below the \T1 decayed qubit excitation at the measurement time $t_{o_f}$. Explicitly, we must have $P_t < e^{-t_{o_f}/\T1}$. The qubit state preparation and measurement visibility affect the maximum qubit excitation that can be measured. The threshold must therefore be chosen below this value. We also need to account for the possibility that some of the $n_s$ samples measured within each bin do not fall exactly at the oscillation minimum. Thus, even is there is a TLS visible in a particular bin, it is likely that $\overline{P_e} > 0$. Consequently, we cannot select an overly low threshold. Finally, since the measurement is a binomial process, we must account for the variance in~$P_e$.

In order to cope with these issues, we determine a threshold by inspecting the average excitation of each bin in Fig.~\ref{fig:octaves}~(b). We find the bin with the highest average excitation value, in our experiment~$\max(\overline{P_e}) = 0.976$. This quantity is a good estimate for the visibility. We then subtract an empirically selected buffer value of~$0.3$ to account for the other effects mentioned above and obtain the threshold~$P_t = 0.676$.

After setting~$P_t$, the average excitation $\overline{P_e}$ for each bin must be compared to it. If for a certain bin~$\overline{P_e} < P_t$, we have detected a resonance mode.

Before estimating the coupling parameters, however, we need to verify whether there are multiple resonance modes within a frequency-time scan or not. This task is achieved by considering all bins meeting the condition~$\overline{P_e} < P_t$ and grouping neighboring bins. If multiple such groups exist, then there are multiple resonance modes interacting with the qubit. In this case, we must split the octave data into multiple frequency-time sub-scans such that there is only one resonance mode per scan. This procedure is explained in Appendix~\ref{ap:octaveanalysis}. Once the data is split by resonance mode, the parameter estimates are given by searching for the below-threshold bin at the \emph{lowest} octave.

We performed an experiment where a qubit interacts with three resonance modes, two of which are synthesized with a microwave source and one is an actual resonator. As expected, our estimation method detects three resonance modes. Their estimated frequency and coupling parameters are reported in Table~\ref{tbl:octaveestimate}. We purposely choose two of the resonance modes to be close in frequency to illustrate an important feature of our estimation method: that two distinct resonance modes are detected as independent modes only if their frequency separation is at least twice as large as the largest coupling strength and twice as large as the minimum frequency width of the bins. The resonator is characterized by a strong coupling coefficient (see Appendix~\ref{ap:setup} for the coupling capacitance). For all three modes, the octave algorithm estimates the frequency properly, though the precision is limited by the bin size. The estimated coupling strength is given by a range. Obtaining more precise and accurate results necessitates the online estimation algorithm to be explained in the next section.

\begin{table}[htb]
  \centering
  \caption{Resonances detected after analysis of the octave sampling data in Fig.~\ref{fig:octaves}~(b).}
  \label{tbl:octavedetection}
  \begin{ruledtabular}
  \begin{tabular}[t]{l r r r}
    Parameter & RM1 & RM2 & RM3 \\
    \hline
    Octave number              &     6 &     7 &       3 \\
    Bin center frequency (GHz) & 4.810 & 4.830 &   5.106 \\
    Bin coupling range  (MHz)  & [4,8] & [2,4] & [32,64] \\
  \end{tabular}
  \end{ruledtabular}
\end{table}

\section{Online Bayesian Learning Algorithm}
\label{sec:bayesianlearning}

The offline octave sampling algorithm is data efficient and can be performed from a state of zero knowledge of the qubit's spectrum. However, it does not provide a very accurate estimate of the coupling parameters. To improve accuracy, we can use the rough estimate given by the offline method to execute an \emph{online Bayesian learning algorithm} and refine the parameters in a very short time.

Given an initial probability distribution over the coupling parameters with a resonance mode, the online algorithm successively selects swap spectroscopy measurement settings to increase knowledge. After the result of a measurement is recorded, the distribution is updated according to Bayes' theorem and a new measurement setting is generated. This procedure is repeated iteratively until the distribution converges as desired~\footnote{The source code developed for this work can be found online at \url{https://gitlab.com/DQMLab/TLSInfer.jl}.}.

\subsection{Theoretical Method}
\label{subs:bayesiantheory}

\begin{figure*}[t]
  \includegraphics{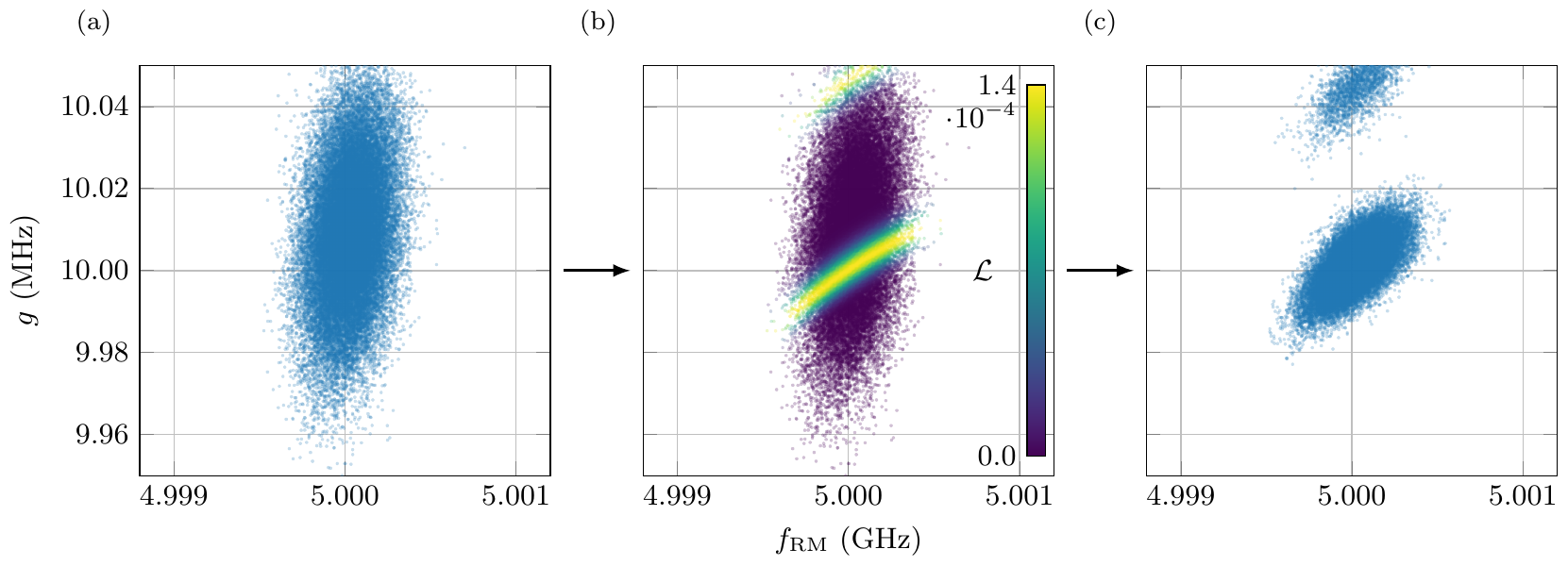}
  \caption{Illustration of a simulated iteration of the online particle filter algorithm with 40000 particles. (a) Given a prior distribution, we heuristically generate measurement settings $f_\text{p}$ and $t$ meant to increase information. (b) Following the measurement, the likelihood~$\mathcal{L}$ of the result is computed for each particle. Note that the values shown on the scale have been normalized. (c) We apply Bayes' theorem to determine the posterior distribution. This task is achieved by resampling the particles according to their likelihoods. The distribution is split in two ``clouds''. After resampling, the posterior distribution can be used as the next iteration's prior.}
  \label{fig:estimationiteration}
\end{figure*}

The online estimation algorithm is an experimental implementation of the one proposed in Ref.~\onlinecite{Stenberg2014}. It employs a particle filter method to efficiently represent the \emph{prior} and \emph{posterior} distributions and compute Bayes' theorem at each iteration.

A particle distribution is a discretized representation of a probability distribution. The denser the distribution in a particular region of the parameter space, the higher the probability of those parameters being valid. Here, each particle represents a value for the coupling parameters~$(f_\text{RM},g)$ of a resonance mode. At the beginning of an iteration, we compute the means, $\mu(f_\text{RM})=\ev{f_\text{RM}}$ and $\mu(g)=\ev{g}$, and standard deviations $\sigma(f_\text{RM})=\sqrt{\ev{f_\text{RM}^2}-\ev{f_\text{RM}}^2}$ and $\sigma(g)=\sqrt{\ev{g^2}-\ev{g}^2}$ of the prior distribution.

The next step is to perform a single swap spectroscopy measurement to determine the excited population~$P_e$ at a particular probe frequency~$f_\text{p}$ and time~$t$. These \emph{measurement settings} are heuristically selected to increase information gain~\cite{Stenberg2014}. In practice, $t$ should scale inversely with~$\sigma(g)$, while $f_\text{p}$ should be within a factor of~$\mu(g)$ on either side of~$\mu(f_\text{RM})$.

We chose the following measurement settings:
\begin{align}
  \label{eq:meassettingsf}
  f_\text{p} &=
  \begin{cases}
    \mu(f_\text{RM}) + r_1 \mu(g) & \text{for $M \leq M_0$,} \\
    \mu(f_\text{RM}) + c r_1 \sigma(f_\text{RM}) & \text{for $M > M_0$,} \\
  \end{cases}
  \\
  \label{eq:meassettingst}
  t &=
  \begin{cases}
    r_2 \tanh[a/\sigma(g) t_\text{max}] t_\text{max} & \text{for $M \leq M_0$,} \\
    \frac{1+r_2}{2} \tanh[a/\sigma(g) t_\text{max}] t_\text{max} & \text{for $M > M_0$,} \\
  \end{cases}
\end{align}
where~$a=\pi/2$, $c=5$, $r_1$ is picked from~$\mathcal{U}(-1/2,1/2)$ and $r_2$ is picked from~$\mathcal{U}(0,1)$. These parameters are empirical constants determined in Ref.~\onlinecite{Stenberg2014}, though they were slightly adjusted for this experiment in order to have a larger distribution for $f_\text{p}$ and $t$ when $M > M_0$. Note that, unlike the method proposed in Ref.~\onlinecite{Stenberg2014}, we choose to limit~$t$ to a maximum value well under~\T1. This is done for the same reason as was explained in Subsec.~\ref{subs:octavetheory}. For this purpose, we use the hyperbolic tangent function as it has a linear behavior for small arguments, such that~$\tanh(a/\sigma(g) t_\text{max}) t_\text{max} \approx a/\sigma(g)$ when~$\sigma(g)$ is large.

After a certain number of iterations~$M_0=25$, we modify the heuristic slightly to accelerate convergence. Initially, we choose probe frequencies coarsely according to the value of~$g$. Then, as our knowledge improves, $\sigma(f_\text{RM})$ decreases and can be used to select frequencies in a narrower range around~$\mu(f_\text{RM})$. The factor~$c$ here is used to avoid choosing measurement frequencies \emph{too} narrowly. The time $t$ is always weighted by~$\sim1/\sigma(g)$, but we bias the selection to larger values after~$M_0$ iterations.

The last step in the iteration is to apply Bayes' theorem to update our knowledge of the coupling parameters. We want to obtain the posterior distribution based on the measurement result~$P_e(f_\text{p},t)$. This is achieved in two sub-steps: first, we compute the likelihood of obtaining the measurement value given each particle's~$(f_\text{RM},g)$ parameters; second, we resample the distribution according to these likelihoods.

We compute the likelihood from the measurement result~$P_e = n_e/n$, which is the proportion of excited state outcomes we recorded~$n_e$ for~$n$ individual measurement shots. Since we know the theoretical fraction that we should expect to measure,~$\tilde{P_e} := \tilde{P_e}(f_\text{p},t,f_\text{RM},g)$, which is given by Eq.~\ref{eq:theoryPe} in the decoherence-free case\footnote{In the experiment we use for~$\tilde{P_e}$ a function that takes into account relaxation as well as measurement visibility. For the relaxation, we use Eq.~(17) in Ref.~\onlinecite{Stenberg2014}. To account for measurement visibility, we clamp the theoretical probability between~$0.05$ and $0.95$.}, we know that the result is a binomial random variable~$n_e \sim B(n,\tilde{P_e})$. Accordingly, the likelihood of obtaining a particular measurement result given measurement settings~$(f_\text{p},t)$ and a particle~$(f_\text{RM},g)$ is the probability mass function
\begin{align}\label{eq:likelihood}
  \mathcal{L}(n_e|f_\text{p},t,f_\text{RM},g) = \binom{n}{n_e} \tilde{P_e}^{n_e} (1-\tilde{P_e})^{(n-n_e)},
\end{align}
where $\binom{n}{n_e}$ is the binomial coefficient.

In effect, we are computing the probability that the measurement result obtained corresponds to a resonance mode with coupling parameters $(f_\text{RM},g)$. The next step is to resample the distribution to keep only those parameters that are most probable. Although this task can be achieved in a variety of ways, the general idea is to pick particles from the prior at random, weighted by the likelihood. To avoid duplicate particles in the posterior distribution, we add normally distributed random noise proportional to the covariance of the prior. The exact procedure used in this work is described in Appendix~\ref{ap:resampling}.

The iteration process is visualized in Fig.~\ref{fig:estimationiteration}, allowing us to understand more intuitively how the particle filter technique works. If the measurement is useful, i.e., the resulting likelihood favors a subset of the prior, the posterior distribution is shrunk or filtered, improving knowledge of the parameters. Otherwise, if the likelihood does not discriminate the particles, the distribution is not modified significantly. After resampling, the next iteration starts with the last iteration's posterior as prior.

Compared to octave sampling, the task of the online Bayesian learning algorithm is simpler since we must already have knowledge of the presence of an interaction. The particle filter can therefore fit to the parameters that are the most likely given the measurements. At the end of the final iteration, the parameters and their uncertainties are given by the mean and standard deviation of the particle distribution. If the algorithm converges, the final particle ``cloud'' is small, resulting in an estimate that is accurate: both true and precise. If the algorithm does not converge, meaning that the final particle cloud is not tightly concentrated in a single region, running the experiment again might be necessary. Regenerating the initial particle distribution via the octave sampling method could also improve detection performance.

Note that the algorithm does not take into account any potential errors on the value of the measurement settings. Therefore, directly taking the standard deviation of the distribution to be the uncertainty on the estimated parameters is valid only in the case where there is no uncertainty on the measurement settings. This will generally not be the case in an experimental setting.

\subsection{Experimental Results}
\label{subs:bayesianresults}

We run the online Bayesian learning algorithm on the three distributions generated from the octave data for each detected resonance mode. The generation of those distributions is discussed in Appendix~\ref{ap:octaveanalysis}. For each mode, we perform 35 iterations of the algorithm, at which point the distribution has converged. The runtime of the algorithm for a single resonance mode is approximately \SI{23}{\second}. We then compute the mean of the distribution after the final iteration. The results are shown in Table~\ref{tbl:onlineres}. As expected, the parameters of the synthesized modes and of the measurement resonator are correctly identified. Standard errors are not shown because the standard deviation of the particle distribution after the final iteration converges to a small value that is not representative of the true error. The error must instead be estimated according to other potential limitations. For our experiment, this should be done according to the characteristics of the flux pulse used to set $f_\text{p}$ and $t$. More details can be found in Appendix~\ref{ap:fluxpulse}.

\begin{table}[ht]
  \centering
  \caption{Estimated parameters for the three resonance modes detected after running the online Bayesian learning algorithm. RM1 and RM2 are synthesized modes and RM3 corresponds to the qubit's measurement resonator.}
  \label{tbl:onlineres}
  \begin{ruledtabular}
  \begin{tabular}[t]{l r r r}
    Parameter           & RM1    & RM2    & RM3    \\
    \hline
    $f_\text{RM}$ (GHz) & 4.8114 & 4.8296 & 5.0860 \\
    $g$ (MHz)           &  3.352 &  1.672 & 43.295 \\
  \end{tabular}
  \end{ruledtabular}
\end{table}

\begin{figure*}[t]
  \includegraphics{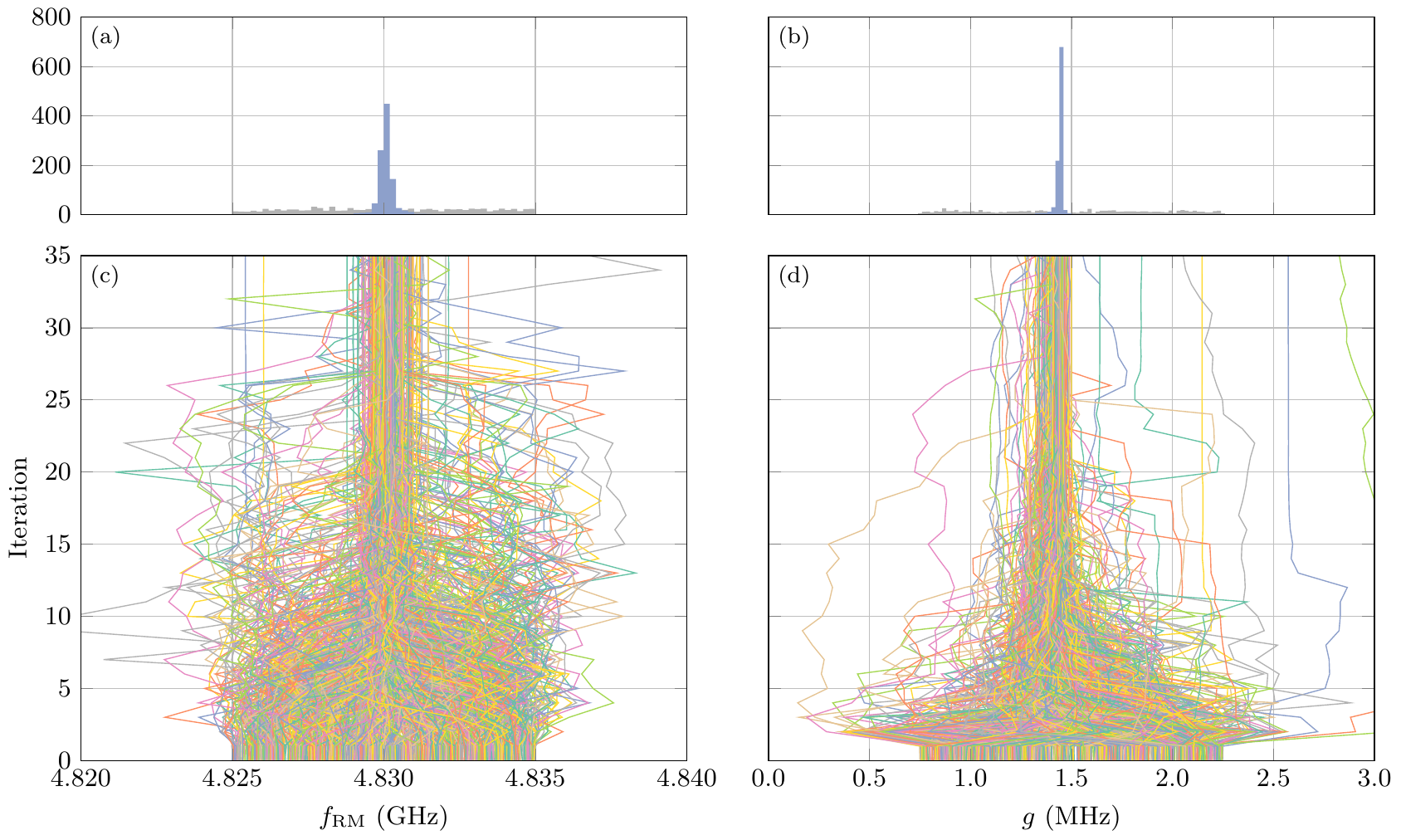}
  \caption{Performance of the online estimation algorithm over~$1000$ runs. The total runtime of the experiment is~\SI{6.4}{\hour}. Each individual run executes~$35$ iterations and, thus, $35$ measurements, taking~$\approx\SI{23}{\second}$. Most of the time (60\%) is spent acquiring data. In fact, each measurement comprises~$786$ shots at a repetition rate of~\SI{2}{\kilo\hertz}. Most of the leftover time is due to data transfer and processing. The total computation time for the estimation algorithm is~$\approx\SI{1}{\second}$.}
  \label{fig:tlsestimation}
\end{figure*}


To test the performance of the estimation algorithm, we run it~$1000$ times on the second resonance mode~(RM2) with slightly different initial distributions. We plot the convergence of the parameters in Fig.~\ref{fig:tlsestimation}. As shown by the histograms, more than~99\% of the runs converge successfully to properly estimate the frequency and coupling strength, with just a few failures. The average of the parameters after the~35\textsuperscript{th} iteration is~$f_\text{RM} = \SI{4.8301(4)}{\giga\hertz}$ and $g = \SI{1.45(9)}{\mega\hertz}$.

We note that one possible cause of failure is the overestimation of~$g$ by an integer multiple; under these conditions, in fact, crests in the oscillations partially overlap. Other possible causes include variance in the parameters of the qubit itself (our probe). If the frequency of the qubit changes abruptly, for example due to appearance of a semi-resonantly coupled TLS defect, the flux-to-frequency calibration becomes invalid; this results in inaccurate measurement settings~\cite{Klimov2018,Burnett2019,Schlor2019}.

The mean of each of the $1000$ initial particle distributions is distributed uniformly at random within a~$\SI{10}{\mega\hertz}$ and $\SI{1.5}{\mega\hertz}$ range for~$f_\text{RM}$ and $g$, respectively. Each individual particle distribution is uniform, with a width of~$\SI{15}{\mega\hertz}$ in~$f_\text{RM}$ and $\SI{2.5}{\mega\hertz}$ in~$g$.

\section{Discussion}
\label{sec:discussion}

Both algorithms presented above depend on a few parameters that are crucial to their function. For the octave sampling algorithm, the choice of the frequency range to be measured is naturally determined by the properties of the device: superconducting qubits have a limited frequency range within which they operate optimally. For the device in this work, the upper end of the measurement range $f_\text{max}$ simply corresponds to the maximum attainable frequency. The lower limit $f_\text{min}$ is chosen to be as low as is possible while still being able to control and measure the qubit. Since we use a resonator for readout, the farther away the qubit is in frequency from the resonator, the lower the fidelity of the measurement. Hence, a qubit cannot operate far away from its readout resonator. Other constraints, e.g., pulse control bandwidth, might dictate even tighter limits.

A second important parameter for the octave sampling method is the number of measurements made per bin $n_s$. In principle, a single high-quality (many shots) measurement of $P_e$ at the center of the bin should be enough. This would also make for a fair comparison with the traditional swap spectroscopy technique. However, because of the efficiency of the octave method, we can afford to take a few more measurements per bin. This is what we have chosen to do, by randomly distributing $n_s=5$ measurements per bin. Another way to take more than one sample could be to sample within bins regularly. In either case, this redundancy serves both to marginally increase the detection sensitivity (by increasing the resolution), or to protect against possible statistical fluctuations in the measurement.


The last parameter we discuss is the number of octaves to be measured, which corresponds to the maximum time $t_{o_f}$, or equivalently, the final frequency resolution $\Delta f_{o_f}$, as explained in Sec.~\ref{subs:octavetheory}. Generally, this parameter should be determined by the requirements of the experiment for which the calibration is made. If a long gate sequence is needed, e.g., for randomized benchmarking, detecting weakly coupled RMs is important. This would not necessarily be the case for shorter experiments, like process tomography. A frequency resolution $\Delta f_{o_f} \sim \frac{1}{t_\text{exp}}$, where $t_\text{exp}$ is the length of the experimental gate sequence, is therefore generally a good choice.

For the particle filter algorithm, the choice of $a$, $c$, and $M_0$ is discussed in Ref.~\onlinecite{Stenberg2014}. Other parameters of interest include $t_\text{max}$, and the number of particles to be used. The time $t_\text{max}$ is used in Eq.~\ref{eq:meassettingst} to restrict the maximum measurement time $t$. This is necessary because the qubit eventually decays to the ground state. To obtain reliable results, $t_\text{max}$ should be set well below \T1. Note that another way to limit the maximum measurement time would be to replace $\tanh(a/\sigma(g) t_\text{max}) t_\text{max}$ with $a/\sigma(g)$ in Eq.~\ref{eq:meassettingst} (as in the original proposal), and simply stop the algorithm once a sufficiently small~$\sigma(g)$ is reached.

The number of particles to be used is constrained mainly by the performance of the computer running the resampling procedure and, potentially, numerical accuracy issues~\cite{Murray2016}. As a rule of thumb, at least 10000 particles should be used; in this work, we have used 40000.

In Sec.~\ref{subs:octavetheory}, we explained that our method can be used not only for a simple qubit swap spectroscopy experiment, but also with a double-excitation protocol, to look for resonances with the $\ket{\text{e}} \leftrightarrow \ket{\text{f}}$ transition. The algorithms discussed in this work are, in fact, more general than that. As long as the qubit dynamics look like a chevron pattern, we can use the same methods to efficiently detect the location and estimate the parameters of the interaction. This is the case, for example, with a whole class of parametric two-qubit gates, where instead of varying the probe frequency $f_\text{p}$ of the qubit, we vary the frequency of a radio-frequency flux drive applied to the DC SQUID of a qubit or tunable coupler~\cite{Bertet2006,McKay2016,Caldwell2018}.

Finally, we briefly discuss the problem of \emph{choosing} qubit operating frequencies. Indeed, once the calibration showcased in this work has been accomplished and all resonant couplings have been identified, the goal is then to make use of the information to optimize performance of a quantum processor. This process is architecture dependent. For an array of directly-coupled superconducting qubits, we want to avoid crosstalk between neighbor qubits, and minimize interactions with TLS defects. We therefore need to choose the idle frequencies of all qubits at the same time, taking into account both wanted and unwanted couplings. Additional concerns apply for choosing the operating frequencies of two-qubit gates: we must consider account the frequency path that the qubits will take during the gate. Having a qubit cross through a resonance with a TLS, for example, is not desirable.

While this work does not explain the process needed to perform this performance optimization (for that see, e.g. Ref.~\onlinecite{Klimov2020}), we emphasize that the runtime improvements of the offline and online algorithms when compared to traditional swap spectroscopy enables several advantages. First, the calibration may be run more often. Second, the calibration is affordable enough to be run on a larger spectrum, and so potentially more resonance modes, giving the frequency optimization process more information to work with.

\section{Conclusion}
\label{sec:conclusion}

In conclusion, we explain two methods for the Hamiltonian parameter estimation of resonant couplings in the context of tunable superconducting qubits. The octave sampling technique can be run without knowledge about the environment of the qubit, and allows efficient detection of coupled resonance modes. The parameter estimation algorithm can be performed or omitted depending on whether a more accurate knowledge of the coupling parameters is desired. Using these algorithms reduces the number of measurements needed by a square-root factor when compared to the traditional method. This translates to a reduction in runtime by one order of magnitude in typical conditions.

We experimentally demonstrate both techniques on an Xmon transmon superconducting qubit and evaluate their performance. We are able to detect the resonance with the qubit's measurement resonator, as well as with synthesized resonance modes. Overall, we determine that the methods are efficient, reliable, and readily automated. We expect this type of calibration to be critical to the operation of large-scale quantum computers, superconducting and not. Future work includes integrating the information we acquire by our methods into a comprehensive optimization process for selecting the operating frequency of each qubit in a quantum processor.

\begin{acknowledgments}
This research was undertaken thanks in part to funding from the Canada First Research Excellence Fund (CFREF) as well as the Discovery and Research Tools and Instruments Grant Programs of the Natural Sciences and Engineering Research Council of Canada (NSERC). The device used was fabricated at the Quantum-Nano Fabrication and Characterization Facility at the University of Waterloo, Canada. We would like to acknowledge the Canadian Microelectronics Corporation (CMC) Microsystems for the provision of products and services that facilitated this research, including CAD tools and design methodology. The authors thank Christopher Warren for useful discussions.
\end{acknowledgments}

\appendix

\section{Setup and Sample}\label{ap:setup}

\begin{figure}[ht]
  \includegraphics{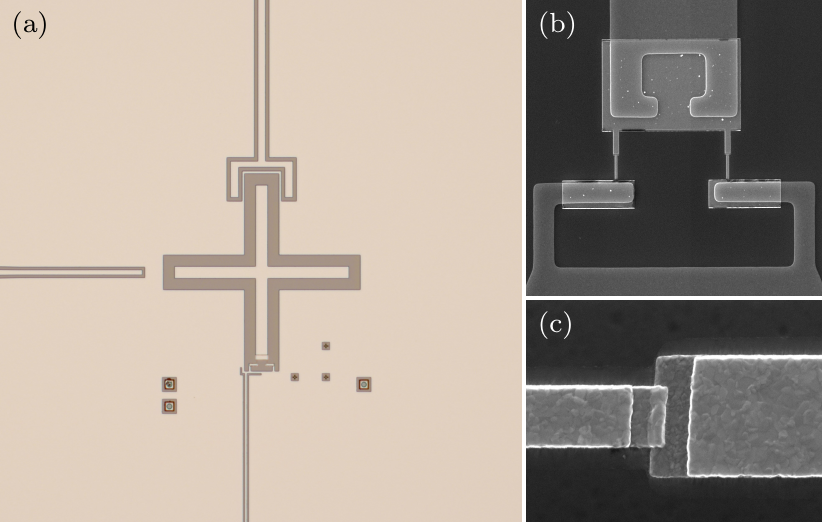}
  \caption{Images of an identically fabricated sample. (a) Optical image of the Xmon transmon qubit, with the drive line on the left, measurement resonator coupler above, and flux bias line below. (b) Scanning electron microscope (SEM) image of the DC SQUID. (c) SEM image of a junction made with a Dolan bridge.}
  \label{fig:sample}
\end{figure}

The sample is a chip comprising two independent Xmon transmon superconducting qubits which are fabricated with evaporated aluminum on a silicon wafer~\cite{Earnest2018}. We use the qubit on the left side of the chip for the experiments shown in this work. Qubit rotations about an axis in the $XY$-plane are controlled by a capacitive microwave drive, while flux pulses to set the frequency are done via a low-frequency line inductively coupled to the qubit's DC SQUID. Measurement is performed by means of a high-power readout scheme using a resonator capacitively coupled to the qubit. Images of an identically-fabricated chip can be seen in Fig.~\ref{fig:sample}. The package-to-chip connections are made with the quantum socket, a fully-vertical packaging method~\cite{Bejanin2016}.

The measurement schematics of the dilution refrigerator, including instrument and wiring details can be found in Fig.~\ref{fig:fridgewiring}. Note that, compared to our previous setup in Ref.~\onlinecite{Bejanin2016}, carbon nanotube-based low-pass filters (LPF) are placed on the microwave lines to prevent infrared noise~\cite{Moghaddam2019}. In addition, coaxial lines were installed for flux biasing. Two lines are used to bias the qubit. One is meant for pure DC current coming from a battery, while the other is used for flux pulses made with an arbitrary waveform generator (AWG). These lines are joined with a custom-made Eccosorb-based bias tee, which acts as a LPF for infrared shielding~\cite{Santavicca2008}.

The parameters of the device used in the experiment are specified in Table~\ref{tbl:deviceparams}. Both \T1 and \T2 fluctuate over the range of accessible qubit frequencies due to the Purcell effect and flux noise.

\begin{table}[ht]
\centering
\caption{Device parameters for the qubit used in the experiment. $C_\text{c}$ is the coupling capacitance between the qubit and its readout resonator. The Hamiltonian parameters are determined via a fit of the resonator and qubit frequency, and the qubit anharmonicity.}
\label{tbl:deviceparams}
\begin{ruledtabular}
\begin{tabular}[t]{l r}
  Parameter & Value \\
  \hline
  $E_J$ & $\SI{19.614(5)}{\giga\hertz}$ \\
  $E_C$ & $\SI{188.92(5)}{\mega\hertz}$ \\
  $f_\text{res}$ & $\SI{5.04844(3)}{\giga\hertz}$ \\
  $C_\text{c}$ & $\SI{3.371(5)}{\femto\farad}$ \\
  $T_1$ & $10$ to $\SI{25}{\micro\second}$ \\
  $T_2$ & $\SI{400}{\nano\second}$ to $\SI{1}{\micro\second}$\\
\end{tabular}
\end{ruledtabular}
\end{table}

\section{Qubit Flux Pulse}\label{ap:fluxpulse}

\begin{figure*}[p]
  \includegraphics{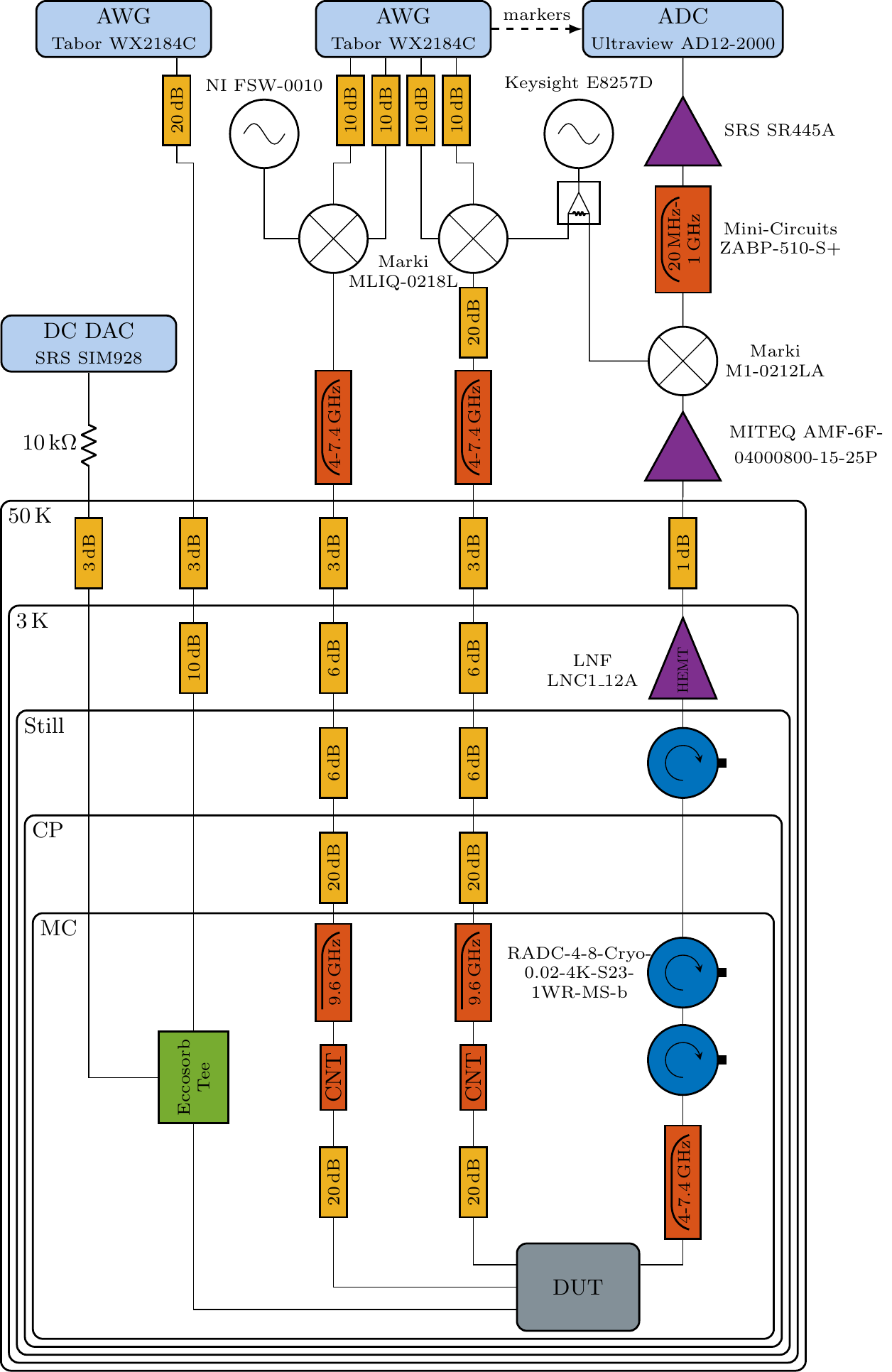}
  \caption{Dilution refrigerator wiring schematics. The sample is mounted in an aluminum package at the mixing chamber stage of a dilution refrigerator, at a temperature of approximately \SI{10}{\milli\kelvin}. We use a dedicated microwave line to drive the qubit. Two coaxial lines are joined at the mixing chamber for the flux biasing of the DC SQUID. We use a homodyne readout setup, with the in-phase quadrature (IQ) mixers configured for image rejection.}
  \label{fig:fridgewiring}
\end{figure*}

Performing swap spectroscopy requires the ability to set the qubit's frequency to a desired value $f_\text{p}$ for a particular duration. In our experiment, this is done with a flux pulse applied to the DC SQUID with an AWG. As shown in Fig.~\ref{fig:fridgewiring}, the flux pulse reaches the sample after going through multiple stages of filtering, attenuation, and connections. This means that the waveform will be modified compared to what is generated by the AWG.

In addition, while we control the \emph{amplitude} of the pulse, we are ultimately interested in the resulting \emph{frequency} of the qubit. We therefore need a way to convert between the amplitude of the flux pulse $A$ and the qubit frequency $f_\text{p}$. This can be done, for example, with pulse spectroscopy, where we send $\pi$-pulses to the qubit at different frequencies while it is detuned by a flux pulse of a particular amplitude. The qubit frequency for that amplitude can then be fit. This is repeated for many amplitudes in order to get a map between $A$ and $f_\text{p}$.

The above considerations mean that the measurement settings $(f_\text{p},t)$ that we select may contain multiple kinds of potential errors. This must be taken into account when estimating the coupling parameters with either the offline or online algorithms.

For example, if $f_\text{p}$ is higher than the true probe frequency of the qubit due to some systematic error, the result $f_\text{RM}$ reported by the online algorithm will be higher than the true value as well. Similarly, an error on the value of $t$ will lead to a wrong estimation of $g$. In practice, this kind of systematic error is not a major problem as long as the error is consistent between experiments.

\section{Details on Octave Analysis}\label{ap:octaveanalysis}

We explain here the method employed to (1) split the raw octave data into spectra containing a single RM, if any, and (2) generate a particle distribution from spectra containing a single RM.

As explained in the main text, we want to make sure that there is only a single resonance in the data when we estimate a frequency and coupling strength. If there are no resonance, or if there is more than one, the particle distribution that we generate is nonsensical. We therefore need to check for bins below the threshold, and split the data accordingly.

To identify resonances, we start from the deepest octave (the one corresponding to the longest time), and look for bins with average excitation below the threshold. If there are multiple consecutive bins below the threshold, we keep only the one with the lowest excitation.

This gives us a list of independent peaks found in the data. We then go to the next deepest octave and perform the same thresholding. At this point it is possible that, if any resonance was found in the previous octave, a bin corresponding to the same resonance is found as well. If that is the case, we keep only the lower octave (shorter time) bin information.

We continue this procedure until the zeroth octave, at which point we are left with the smallest octave (the lowest time) bin for every resonance detected. To split the spectrum, we simply cut the data at the midpoint of each detected resonance. If this cut happens to fall within a bin, we slice the bin in two, making sure to keep track of each part's proportion. For the data in Fig.~\ref{fig:octaves}~(b), we detected three resonances, tabulated in Table~\ref{tbl:octavedetection}.


Following this procedure, we can restrict ourselves to a frequency-time scan containing a single resonance mode. We can specify a discrete distribution representing our knowledge of the frequency and coupling parameters between the qubit and the resonance mode. We thus choose all the bins below threshold and weigh them following a heuristic procedure to find \emph{how much below threshold} they are. This allows us to draw a \emph{particle} distribution associated with the bins as explained in the following.

Each particle represents a frequency---coupling-strength pair. We choose a bin according to its weight and generate a particle within it. In this experiment, the weight of each bin is chosen to be proportional to $\max(\overline{P_e})-\overline{P_e}$. Thus, bins for which the average excitation $\overline{P_e}$ is low have a high weight. The frequency of the generated particle is chosen uniformly at random within the bin bounds and its coupling strength is selected from~$\mathcal{U}\left(\Delta_o,2\Delta_o\right)$, where~$\Delta_o$ is the bin width. We then repeat this procedure to generate a full distribution.

\begin{figure}[htb]
  \includegraphics[width=\linewidth]{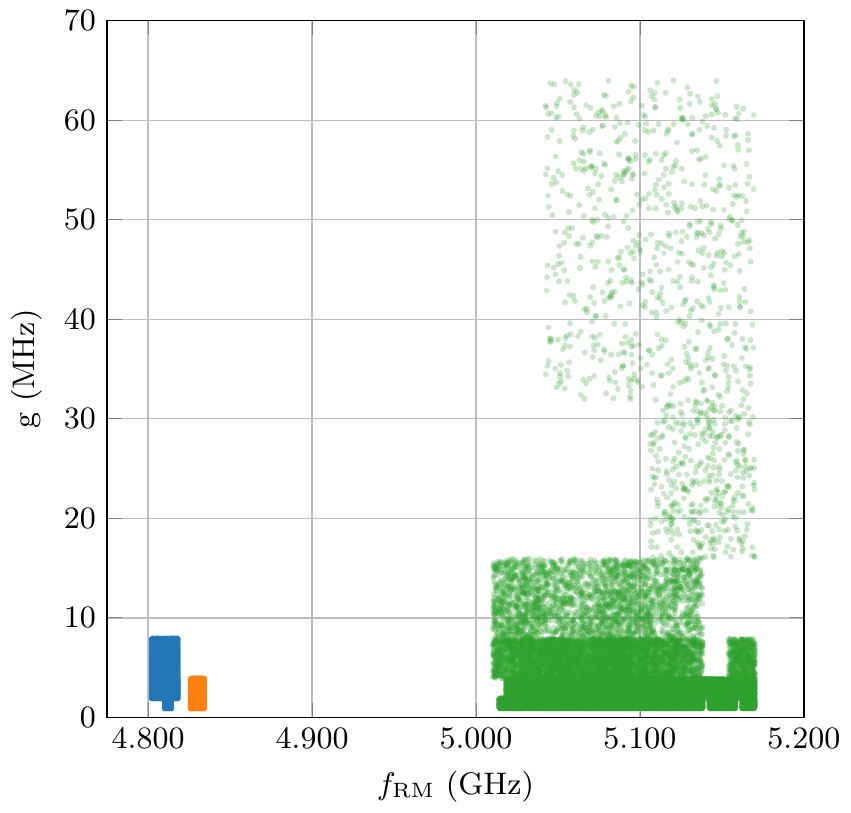}
  \caption{Initial particle distribution generated from the octave data in Fig.~\ref{fig:octaves}~(b) after splitting the spectrum in three parts.}
  \label{fig:initparticles}
\end{figure}

The three resulting distributions can be seen in Fig.~\ref{fig:initparticles}, with their statistics tabulated in Table~\ref{tbl:octaveestimate}. We use a distribution comprising 40000 particles, though this number can be adjusted depending on the capabilities of the computer controlling the experiment. These particle distributions can be used as the starting point for the online algorithm discussed in Sec.~\ref{sec:bayesianlearning}.

\begin{table}[htb]
  \centering
  \caption{Frequency and coupling parameters for three resonance modes computed from the data in Fig.~\ref{fig:octaves}~(b) which was obtained with the offline octave sampling algorithm.}
  \label{tbl:octaveestimate}
  \begin{ruledtabular}
  \begin{tabular}[t]{l r r r}
    Parameter & RM1 & RM2 & RM3 \\
    \hline
    $f_\text{RM}$ (GHz) & 4.811(4) & 4.830(2) & 5.088(42) \\
    $g$ (MHz)           &     3(2) &     2(1) &      4(7) \\
  \end{tabular}
  \end{ruledtabular}
\end{table}

\section{Bayesian Resampling Procedure}\label{ap:resampling}

The exact resampling procedure is described in detail in Chap.~2 of Ref.~\onlinecite{Sanders2016}. We reproduce a shortened version here for completeness.

As inputs, we require the particle locations, as an array of frequency---coupling-strength pairs, in addition to their likelihoods, as computed in the main article. Note that the likelihoods must be normalized to sum to unity, at which point they become \emph{weights}. This normalization ensures that the weights array is a valid discrete probability distribution that can be drawn from.

The algorithm then draws particles from the input distribution according to the weights, and, from the position of the particles drawn, generate new ones by adding ``noise''. This prevents having duplicates in the output distribution, even if a particle from the input is drawn multiple times. The resampling algorithm is shown in pseudocode in Alg.~\ref{alg:resampling}.

\begin{algorithm}[H]\label{alg:resampling}
  \DontPrintSemicolon
  \SetKwProg{Fn}{Function}{}{end}
  \KwIn{Array of particle positions $\{\vec{x}_k\}$}
  \KwIn{Array of particle weights $\{w_k\}$}
  \KwOut{New particle positions $\{\vec{y}_k\}$}
  \Fn{resample($\{\vec{x}_k\}$,$\{w_k\}$)}{
   $a = 0.98$\;
   $\vec{\mu} = \text{mean}(\{\vec{x}_k\})$\;
   $\vec{\Sigma} = (1-a^2) \text{cov}(\{\vec{x}_k\})$\;
   \For{$i \in 1:n$}{
    $l = \text{rand}(\text{Discrete}(\{w_k\}))$ \;
    $\vec{\mu}_l = a\vec{x}_l + (1-a)\vec{\mu}$\;
    $\vec{y}_i = \text{rand}(\text{Normal}(\vec{\mu}_l,\vec{\Sigma}))$\;
    }
  \Return{$\{\vec{y}_k\}$}
  }
  \caption{Particle Resampling}
\end{algorithm}

Note that the particle at index $i$ in the $\{\vec{x}_k\}$ array has its corresponding weight at the same index in the $\{w_k\}$ array. In addition, the amount of noise added to the position of the particle drawn is controlled by the spread of the input distribution, quantified by taking the covariance. For more details, see Ref.~\onlinecite{Sanders2016}.

\begin{thebibliography}{52}%
\makeatletter
\providecommand \@ifxundefined [1]{%
 \@ifx{#1\undefined}
}%
\providecommand \@ifnum [1]{%
 \ifnum #1\expandafter \@firstoftwo
 \else \expandafter \@secondoftwo
 \fi
}%
\providecommand \@ifx [1]{%
 \ifx #1\expandafter \@firstoftwo
 \else \expandafter \@secondoftwo
 \fi
}%
\providecommand \natexlab [1]{#1}%
\providecommand \enquote  [1]{``#1''}%
\providecommand \bibnamefont  [1]{#1}%
\providecommand \bibfnamefont [1]{#1}%
\providecommand \citenamefont [1]{#1}%
\providecommand \href@noop [0]{\@secondoftwo}%
\providecommand \href [0]{\begingroup \@sanitize@url \@href}%
\providecommand \@href[1]{\@@startlink{#1}\@@href}%
\providecommand \@@href[1]{\endgroup#1\@@endlink}%
\providecommand \@sanitize@url [0]{\catcode `\\12\catcode `\$12\catcode
  `\&12\catcode `\#12\catcode `\^12\catcode `\_12\catcode `\%12\relax}%
\providecommand \@@startlink[1]{}%
\providecommand \@@endlink[0]{}%
\providecommand \url  [0]{\begingroup\@sanitize@url \@url }%
\providecommand \@url [1]{\endgroup\@href {#1}{\urlprefix }}%
\providecommand \urlprefix  [0]{URL }%
\providecommand \Eprint [0]{\href }%
\providecommand \doibase [0]{https://doi.org/}%
\providecommand \selectlanguage [0]{\@gobble}%
\providecommand \bibinfo  [0]{\@secondoftwo}%
\providecommand \bibfield  [0]{\@secondoftwo}%
\providecommand \translation [1]{[#1]}%
\providecommand \BibitemOpen [0]{}%
\providecommand \bibitemStop [0]{}%
\providecommand \bibitemNoStop [0]{.\EOS\space}%
\providecommand \EOS [0]{\spacefactor3000\relax}%
\providecommand \BibitemShut  [1]{\csname bibitem#1\endcsname}%
\let\auto@bib@innerbib\@empty
\bibitem [{\citenamefont {Ladd}\ \emph {et~al.}(2010)\citenamefont {Ladd},
  \citenamefont {Jelezko}, \citenamefont {Laflamme}, \citenamefont {Nakamura},
  \citenamefont {Monroe},\ and\ \citenamefont {O'Brien}}]{Ladd2010}%
  \BibitemOpen
  \bibfield  {author} {\bibinfo {author} {\bibfnamefont {T.~D.}\ \bibnamefont
  {Ladd}}, \bibinfo {author} {\bibfnamefont {F.}~\bibnamefont {Jelezko}},
  \bibinfo {author} {\bibfnamefont {R.}~\bibnamefont {Laflamme}}, \bibinfo
  {author} {\bibfnamefont {Y.}~\bibnamefont {Nakamura}}, \bibinfo {author}
  {\bibfnamefont {C.}~\bibnamefont {Monroe}},\ and\ \bibinfo {author}
  {\bibfnamefont {J.~L.}\ \bibnamefont {O'Brien}},\ }\bibfield  {title}
  {\bibinfo {title} {{Quantum computers}},\ }\href
  {https://doi.org/10.1038/nature08812} {\bibfield  {journal} {\bibinfo
  {journal} {Nature}\ }\textbf {\bibinfo {volume} {464}},\ \bibinfo {pages}
  {45} (\bibinfo {year} {2010})}\BibitemShut {NoStop}%
\bibitem [{\citenamefont {Gottesman}(2010)}]{Gottesman2010}%
  \BibitemOpen
  \bibfield  {author} {\bibinfo {author} {\bibfnamefont {D.}~\bibnamefont
  {Gottesman}},\ }\bibfield  {title} {\bibinfo {title} {{An introduction to
  quantum error correction and fault-tolerant quantum computation}},\ }in\
  \href {https://doi.org/10.1090/psapm/068/2762145} {\emph {\bibinfo
  {booktitle} {Quantum Information Science and Its Contributions to
  Mathematics}}}\ (\bibinfo  {publisher} {American Mathematical Society},\
  \bibinfo {year} {2010})\ pp.\ \bibinfo {pages} {13--58}\BibitemShut {NoStop}%
\bibitem [{\citenamefont {Fowler}\ \emph {et~al.}(2012)\citenamefont {Fowler},
  \citenamefont {Mariantoni}, \citenamefont {Martinis},\ and\ \citenamefont
  {Cleland}}]{Fowler2012}%
  \BibitemOpen
  \bibfield  {author} {\bibinfo {author} {\bibfnamefont {A.~G.}\ \bibnamefont
  {Fowler}}, \bibinfo {author} {\bibfnamefont {M.}~\bibnamefont {Mariantoni}},
  \bibinfo {author} {\bibfnamefont {J.~M.}\ \bibnamefont {Martinis}},\ and\
  \bibinfo {author} {\bibfnamefont {A.~N.}\ \bibnamefont {Cleland}},\
  }\bibfield  {title} {\bibinfo {title} {{Surface codes: Towards practical
  large-scale quantum computation}},\ }\href
  {https://doi.org/10.1103/PhysRevA.86.032324} {\bibfield  {journal} {\bibinfo
  {journal} {Physical Review A}\ }\textbf {\bibinfo {volume} {86}},\ \bibinfo
  {pages} {032324} (\bibinfo {year} {2012})}\BibitemShut {NoStop}%
\bibitem [{\citenamefont {Kechedzhi}\ \emph {et~al.}(2018)\citenamefont
  {Kechedzhi}, \citenamefont {Smelyanskiy}, \citenamefont {McClean},
  \citenamefont {Denchev}, \citenamefont {Mohseni}, \citenamefont {Isakov},
  \citenamefont {Boixo}, \citenamefont {Altshuler},\ and\ \citenamefont
  {Neven}}]{Kechedzhi2018}%
  \BibitemOpen
  \bibfield  {author} {\bibinfo {author} {\bibfnamefont {K.}~\bibnamefont
  {Kechedzhi}}, \bibinfo {author} {\bibfnamefont {V.}~\bibnamefont
  {Smelyanskiy}}, \bibinfo {author} {\bibfnamefont {J.~R.}\ \bibnamefont
  {McClean}}, \bibinfo {author} {\bibfnamefont {V.~S.}\ \bibnamefont
  {Denchev}}, \bibinfo {author} {\bibfnamefont {M.}~\bibnamefont {Mohseni}},
  \bibinfo {author} {\bibfnamefont {S.}~\bibnamefont {Isakov}}, \bibinfo
  {author} {\bibfnamefont {S.}~\bibnamefont {Boixo}}, \bibinfo {author}
  {\bibfnamefont {B.}~\bibnamefont {Altshuler}},\ and\ \bibinfo {author}
  {\bibfnamefont {H.}~\bibnamefont {Neven}},\ }\bibfield  {title} {\bibinfo
  {title} {{Efficient Population Transfer via Non-Ergodic Extended States in
  Quantum Spin Glass}},\ }in\ \href {https://doi.org/10.4230/LIPIcs.TQC.2018.9}
  {\emph {\bibinfo {booktitle} {Leibniz International Proceedings in
  Informatics (LIPIcs)}}},\ Vol.\ \bibinfo {volume} {111},\ \bibinfo {editor}
  {edited by\ \bibinfo {editor} {\bibfnamefont {S.}~\bibnamefont {Jeffery}}}\
  (\bibinfo  {publisher} {Schloss Dagstuhl},\ \bibinfo {address} {Dagstuhl,
  Germany},\ \bibinfo {year} {2018})\ pp.\ \bibinfo {pages}
  {9:1--9:16}\BibitemShut {NoStop}%
\bibitem [{\citenamefont {Somma}\ \emph {et~al.}(2008)\citenamefont {Somma},
  \citenamefont {Boixo}, \citenamefont {Barnum},\ and\ \citenamefont
  {Knill}}]{Somma2008}%
  \BibitemOpen
  \bibfield  {author} {\bibinfo {author} {\bibfnamefont {R.~D.}\ \bibnamefont
  {Somma}}, \bibinfo {author} {\bibfnamefont {S.}~\bibnamefont {Boixo}},
  \bibinfo {author} {\bibfnamefont {H.}~\bibnamefont {Barnum}},\ and\ \bibinfo
  {author} {\bibfnamefont {E.}~\bibnamefont {Knill}},\ }\bibfield  {title}
  {\bibinfo {title} {{Quantum Simulations of Classical Annealing Processes}},\
  }\href {https://doi.org/10.1103/PhysRevLett.101.130504} {\bibfield  {journal}
  {\bibinfo  {journal} {Physical Review Letters}\ }\textbf {\bibinfo {volume}
  {101}},\ \bibinfo {pages} {130504} (\bibinfo {year} {2008})}\BibitemShut
  {NoStop}%
\bibitem [{\citenamefont {Farhi}\ and\ \citenamefont
  {Neven}(2018)}]{Farhi2018}%
  \BibitemOpen
  \bibfield  {author} {\bibinfo {author} {\bibfnamefont {E.}~\bibnamefont
  {Farhi}}\ and\ \bibinfo {author} {\bibfnamefont {H.}~\bibnamefont {Neven}},\
  }\bibfield  {title} {\bibinfo {title} {{Classification with Quantum Neural
  Networks on Near Term Processors}},\ }\Eprint
  {https://arxiv.org/abs/1802.06002} {arXiv:1802.06002}  (\bibinfo {year}
  {2018})\BibitemShut {NoStop}%
\bibitem [{\citenamefont {McClean}\ \emph {et~al.}(2018)\citenamefont
  {McClean}, \citenamefont {Boixo}, \citenamefont {Smelyanskiy}, \citenamefont
  {Babbush},\ and\ \citenamefont {Neven}}]{McClean2018}%
  \BibitemOpen
  \bibfield  {author} {\bibinfo {author} {\bibfnamefont {J.~R.}\ \bibnamefont
  {McClean}}, \bibinfo {author} {\bibfnamefont {S.}~\bibnamefont {Boixo}},
  \bibinfo {author} {\bibfnamefont {V.~N.}\ \bibnamefont {Smelyanskiy}},
  \bibinfo {author} {\bibfnamefont {R.}~\bibnamefont {Babbush}},\ and\ \bibinfo
  {author} {\bibfnamefont {H.}~\bibnamefont {Neven}},\ }\bibfield  {title}
  {\bibinfo {title} {{Barren plateaus in quantum neural network training
  landscapes}},\ }\href {https://doi.org/10.1038/s41467-018-07090-4} {\bibfield
   {journal} {\bibinfo  {journal} {Nature Communications}\ }\textbf {\bibinfo
  {volume} {9}},\ \bibinfo {pages} {4812} (\bibinfo {year} {2018})}\BibitemShut
  {NoStop}%
\bibitem [{\citenamefont {Cong}\ \emph {et~al.}(2019)\citenamefont {Cong},
  \citenamefont {Choi},\ and\ \citenamefont {Lukin}}]{Cong2019}%
  \BibitemOpen
  \bibfield  {author} {\bibinfo {author} {\bibfnamefont {I.}~\bibnamefont
  {Cong}}, \bibinfo {author} {\bibfnamefont {S.}~\bibnamefont {Choi}},\ and\
  \bibinfo {author} {\bibfnamefont {M.~D.}\ \bibnamefont {Lukin}},\ }\bibfield
  {title} {\bibinfo {title} {{Quantum convolutional neural networks}},\ }\href
  {https://doi.org/10.1038/s41567-019-0648-8} {\bibfield  {journal} {\bibinfo
  {journal} {Nature Physics}\ }\textbf {\bibinfo {volume} {15}},\ \bibinfo
  {pages} {1273} (\bibinfo {year} {2019})}\BibitemShut {NoStop}%
\bibitem [{\citenamefont {Bravyi}\ \emph {et~al.}(2018)\citenamefont {Bravyi},
  \citenamefont {Gosset},\ and\ \citenamefont {K{\"{o}}nig}}]{Bravyi2018}%
  \BibitemOpen
  \bibfield  {author} {\bibinfo {author} {\bibfnamefont {S.}~\bibnamefont
  {Bravyi}}, \bibinfo {author} {\bibfnamefont {D.}~\bibnamefont {Gosset}},\
  and\ \bibinfo {author} {\bibfnamefont {R.}~\bibnamefont {K{\"{o}}nig}},\
  }\bibfield  {title} {\bibinfo {title} {{Quantum advantage with shallow
  circuits}},\ }\href {https://doi.org/10.1126/science.aar3106} {\bibfield
  {journal} {\bibinfo  {journal} {Science}\ }\textbf {\bibinfo {volume}
  {362}},\ \bibinfo {pages} {308} (\bibinfo {year} {2018})}\BibitemShut
  {NoStop}%
\bibitem [{\citenamefont {Aspuru-Guzik}(2005)}]{Aspuru-Guzik2005}%
  \BibitemOpen
  \bibfield  {author} {\bibinfo {author} {\bibfnamefont {A.}~\bibnamefont
  {Aspuru-Guzik}},\ }\bibfield  {title} {\bibinfo {title} {{Simulated Quantum
  Computation of Molecular Energies}},\ }\href
  {https://doi.org/10.1126/science.1113479} {\bibfield  {journal} {\bibinfo
  {journal} {Science}\ }\textbf {\bibinfo {volume} {309}},\ \bibinfo {pages}
  {1704} (\bibinfo {year} {2005})}\BibitemShut {NoStop}%
\bibitem [{\citenamefont {Peruzzo}\ \emph {et~al.}(2014)\citenamefont
  {Peruzzo}, \citenamefont {McClean}, \citenamefont {Shadbolt}, \citenamefont
  {Yung}, \citenamefont {Zhou}, \citenamefont {Love}, \citenamefont
  {Aspuru-Guzik},\ and\ \citenamefont {O'Brien}}]{Peruzzo2014}%
  \BibitemOpen
  \bibfield  {author} {\bibinfo {author} {\bibfnamefont {A.}~\bibnamefont
  {Peruzzo}}, \bibinfo {author} {\bibfnamefont {J.}~\bibnamefont {McClean}},
  \bibinfo {author} {\bibfnamefont {P.}~\bibnamefont {Shadbolt}}, \bibinfo
  {author} {\bibfnamefont {M.-H.}\ \bibnamefont {Yung}}, \bibinfo {author}
  {\bibfnamefont {X.-Q.}\ \bibnamefont {Zhou}}, \bibinfo {author}
  {\bibfnamefont {P.~J.}\ \bibnamefont {Love}}, \bibinfo {author}
  {\bibfnamefont {A.}~\bibnamefont {Aspuru-Guzik}},\ and\ \bibinfo {author}
  {\bibfnamefont {J.~L.}\ \bibnamefont {O'Brien}},\ }\bibfield  {title}
  {\bibinfo {title} {{A variational eigenvalue solver on a photonic quantum
  processor}},\ }\href {https://doi.org/10.1038/ncomms5213} {\bibfield
  {journal} {\bibinfo  {journal} {Nature Communications}\ }\textbf {\bibinfo
  {volume} {5}},\ \bibinfo {pages} {4213} (\bibinfo {year} {2014})}\BibitemShut
  {NoStop}%
\bibitem [{\citenamefont {Hempel}\ \emph {et~al.}(2018)\citenamefont {Hempel},
  \citenamefont {Maier}, \citenamefont {Romero}, \citenamefont {McClean},
  \citenamefont {Monz}, \citenamefont {Shen}, \citenamefont {Jurcevic},
  \citenamefont {Lanyon}, \citenamefont {Love}, \citenamefont {Babbush},
  \citenamefont {Aspuru-Guzik}, \citenamefont {Blatt},\ and\ \citenamefont
  {Roos}}]{Hempel2018}%
  \BibitemOpen
  \bibfield  {author} {\bibinfo {author} {\bibfnamefont {C.}~\bibnamefont
  {Hempel}}, \bibinfo {author} {\bibfnamefont {C.}~\bibnamefont {Maier}},
  \bibinfo {author} {\bibfnamefont {J.}~\bibnamefont {Romero}}, \bibinfo
  {author} {\bibfnamefont {J.}~\bibnamefont {McClean}}, \bibinfo {author}
  {\bibfnamefont {T.}~\bibnamefont {Monz}}, \bibinfo {author} {\bibfnamefont
  {H.}~\bibnamefont {Shen}}, \bibinfo {author} {\bibfnamefont {P.}~\bibnamefont
  {Jurcevic}}, \bibinfo {author} {\bibfnamefont {B.~P.}\ \bibnamefont
  {Lanyon}}, \bibinfo {author} {\bibfnamefont {P.}~\bibnamefont {Love}},
  \bibinfo {author} {\bibfnamefont {R.}~\bibnamefont {Babbush}}, \bibinfo
  {author} {\bibfnamefont {A.}~\bibnamefont {Aspuru-Guzik}}, \bibinfo {author}
  {\bibfnamefont {R.}~\bibnamefont {Blatt}},\ and\ \bibinfo {author}
  {\bibfnamefont {C.~F.}\ \bibnamefont {Roos}},\ }\bibfield  {title} {\bibinfo
  {title} {{Quantum Chemistry Calculations on a Trapped-Ion Quantum
  Simulator}},\ }\href {https://doi.org/10.1103/PhysRevX.8.031022} {\bibfield
  {journal} {\bibinfo  {journal} {Physical Review X}\ }\textbf {\bibinfo
  {volume} {8}},\ \bibinfo {pages} {031022} (\bibinfo {year}
  {2018})}\BibitemShut {NoStop}%
\bibitem [{\citenamefont {Mohseni}\ \emph {et~al.}(2017)\citenamefont
  {Mohseni}, \citenamefont {Read}, \citenamefont {Neven}, \citenamefont
  {Boixo}, \citenamefont {Denchev}, \citenamefont {Babbush}, \citenamefont
  {Fowler}, \citenamefont {Smelyanskiy},\ and\ \citenamefont
  {Martinis}}]{Mohseni2017}%
  \BibitemOpen
  \bibfield  {author} {\bibinfo {author} {\bibfnamefont {M.}~\bibnamefont
  {Mohseni}}, \bibinfo {author} {\bibfnamefont {P.}~\bibnamefont {Read}},
  \bibinfo {author} {\bibfnamefont {H.}~\bibnamefont {Neven}}, \bibinfo
  {author} {\bibfnamefont {S.}~\bibnamefont {Boixo}}, \bibinfo {author}
  {\bibfnamefont {V.~S.}\ \bibnamefont {Denchev}}, \bibinfo {author}
  {\bibfnamefont {R.}~\bibnamefont {Babbush}}, \bibinfo {author} {\bibfnamefont
  {A.}~\bibnamefont {Fowler}}, \bibinfo {author} {\bibfnamefont
  {V.}~\bibnamefont {Smelyanskiy}},\ and\ \bibinfo {author} {\bibfnamefont
  {J.}~\bibnamefont {Martinis}},\ }\bibfield  {title} {\bibinfo {title}
  {{Commercialize quantum technologies in five years}},\ }\href
  {https://doi.org/10.1038/543171a} {\bibfield  {journal} {\bibinfo  {journal}
  {Nature}\ }\textbf {\bibinfo {volume} {543}},\ \bibinfo {pages} {171}
  (\bibinfo {year} {2017})}\BibitemShut {NoStop}%
\bibitem [{\citenamefont {Clarke}\ and\ \citenamefont
  {Wilhelm}(2008)}]{Clarke2008}%
  \BibitemOpen
  \bibfield  {author} {\bibinfo {author} {\bibfnamefont {J.}~\bibnamefont
  {Clarke}}\ and\ \bibinfo {author} {\bibfnamefont {F.~K.}\ \bibnamefont
  {Wilhelm}},\ }\bibfield  {title} {\bibinfo {title} {{Superconducting quantum
  bits}},\ }\href {https://doi.org/10.1038/nature07128} {\bibfield  {journal}
  {\bibinfo  {journal} {Nature}\ }\textbf {\bibinfo {volume} {453}},\ \bibinfo
  {pages} {1031} (\bibinfo {year} {2008})}\BibitemShut {NoStop}%
\bibitem [{\citenamefont {Wendin}(2017)}]{Wendin2017}%
  \BibitemOpen
  \bibfield  {author} {\bibinfo {author} {\bibfnamefont {G.}~\bibnamefont
  {Wendin}},\ }\bibfield  {title} {\bibinfo {title} {{Quantum information
  processing with superconducting circuits: a review}},\ }\href
  {https://doi.org/10.1088/1361-6633/aa7e1a} {\bibfield  {journal} {\bibinfo
  {journal} {Reports on Progress in Physics}\ }\textbf {\bibinfo {volume}
  {80}},\ \bibinfo {pages} {106001} (\bibinfo {year} {2017})}\BibitemShut
  {NoStop}%
\bibitem [{\citenamefont {Kockum}\ and\ \citenamefont
  {Nori}(2019)}]{Kockum2019}%
  \BibitemOpen
  \bibfield  {author} {\bibinfo {author} {\bibfnamefont {A.~F.}\ \bibnamefont
  {Kockum}}\ and\ \bibinfo {author} {\bibfnamefont {F.}~\bibnamefont {Nori}},\
  }\bibfield  {title} {\bibinfo {title} {{Quantum Bits with Josephson
  Junctions}},\ }in\ \href {https://doi.org/10.1007/978-3-030-20726-7_17}
  {\emph {\bibinfo {booktitle} {Fundamentals and Frontiers of the Josephson
  Effect}}}\ (\bibinfo  {publisher} {Springer},\ \bibinfo {year} {2019})\ pp.\
  \bibinfo {pages} {703--741}\BibitemShut {NoStop}%
\bibitem [{\citenamefont {Mariantoni}(2020)}]{Mariantoni2020}%
  \BibitemOpen
  \bibfield  {author} {\bibinfo {author} {\bibfnamefont {M.}~\bibnamefont
  {Mariantoni}},\ }\bibfield  {title} {\bibinfo {title} {{The Energy of an
  Arbitrary Electrical Circuit, Classical and Quantum}},\ }\Eprint
  {https://arxiv.org/abs/2007.08519} {arXiv:2007.08519}  (\bibinfo {year}
  {2020})\BibitemShut {NoStop}%
\bibitem [{\citenamefont {Mariantoni}\ \emph
  {et~al.}(2011{\natexlab{a}})\citenamefont {Mariantoni}, \citenamefont {Wang},
  \citenamefont {Yamamoto}, \citenamefont {Neeley}, \citenamefont {Bialczak},
  \citenamefont {Chen}, \citenamefont {Lenander}, \citenamefont {Lucero},
  \citenamefont {O'Connell}, \citenamefont {Sank}, \citenamefont {Weides},
  \citenamefont {Wenner}, \citenamefont {Yin}, \citenamefont {Zhao},
  \citenamefont {Korotkov}, \citenamefont {Cleland},\ and\ \citenamefont
  {Martinis}}]{Mariantoni2011}%
  \BibitemOpen
  \bibfield  {author} {\bibinfo {author} {\bibfnamefont {M.}~\bibnamefont
  {Mariantoni}}, \bibinfo {author} {\bibfnamefont {H.}~\bibnamefont {Wang}},
  \bibinfo {author} {\bibfnamefont {T.}~\bibnamefont {Yamamoto}}, \bibinfo
  {author} {\bibfnamefont {M.}~\bibnamefont {Neeley}}, \bibinfo {author}
  {\bibfnamefont {R.~C.}\ \bibnamefont {Bialczak}}, \bibinfo {author}
  {\bibfnamefont {Y.}~\bibnamefont {Chen}}, \bibinfo {author} {\bibfnamefont
  {M.}~\bibnamefont {Lenander}}, \bibinfo {author} {\bibfnamefont
  {E.}~\bibnamefont {Lucero}}, \bibinfo {author} {\bibfnamefont {A.~D.}\
  \bibnamefont {O'Connell}}, \bibinfo {author} {\bibfnamefont {D.}~\bibnamefont
  {Sank}}, \bibinfo {author} {\bibfnamefont {M.}~\bibnamefont {Weides}},
  \bibinfo {author} {\bibfnamefont {J.}~\bibnamefont {Wenner}}, \bibinfo
  {author} {\bibfnamefont {Y.}~\bibnamefont {Yin}}, \bibinfo {author}
  {\bibfnamefont {J.}~\bibnamefont {Zhao}}, \bibinfo {author} {\bibfnamefont
  {a.~N.}\ \bibnamefont {Korotkov}}, \bibinfo {author} {\bibfnamefont {A.~N.}\
  \bibnamefont {Cleland}},\ and\ \bibinfo {author} {\bibfnamefont {J.~M.}\
  \bibnamefont {Martinis}},\ }\bibfield  {title} {\bibinfo {title}
  {{Implementing the quantum von Neumann architecture with superconducting
  circuits.}},\ }\href {https://doi.org/10.1126/science.1208517} {\bibfield
  {journal} {\bibinfo  {journal} {Science Advances}\ }\textbf {\bibinfo
  {volume} {334}},\ \bibinfo {pages} {61} (\bibinfo {year}
  {2011}{\natexlab{a}})}\BibitemShut {NoStop}%
\bibitem [{\citenamefont {Lucero}\ \emph {et~al.}(2012)\citenamefont {Lucero},
  \citenamefont {Barends}, \citenamefont {Chen}, \citenamefont {Kelly},
  \citenamefont {Mariantoni}, \citenamefont {Megrant}, \citenamefont
  {O'Malley}, \citenamefont {Sank}, \citenamefont {Vainsencher}, \citenamefont
  {Wenner}, \citenamefont {White}, \citenamefont {Yin}, \citenamefont
  {Cleland},\ and\ \citenamefont {Martinis}}]{Lucero2012}%
  \BibitemOpen
  \bibfield  {author} {\bibinfo {author} {\bibfnamefont {E.}~\bibnamefont
  {Lucero}}, \bibinfo {author} {\bibfnamefont {R.}~\bibnamefont {Barends}},
  \bibinfo {author} {\bibfnamefont {Y.}~\bibnamefont {Chen}}, \bibinfo {author}
  {\bibfnamefont {J.}~\bibnamefont {Kelly}}, \bibinfo {author} {\bibfnamefont
  {M.}~\bibnamefont {Mariantoni}}, \bibinfo {author} {\bibfnamefont
  {A.}~\bibnamefont {Megrant}}, \bibinfo {author} {\bibfnamefont
  {P.}~\bibnamefont {O'Malley}}, \bibinfo {author} {\bibfnamefont
  {D.}~\bibnamefont {Sank}}, \bibinfo {author} {\bibfnamefont {A.}~\bibnamefont
  {Vainsencher}}, \bibinfo {author} {\bibfnamefont {J.}~\bibnamefont {Wenner}},
  \bibinfo {author} {\bibfnamefont {T.}~\bibnamefont {White}}, \bibinfo
  {author} {\bibfnamefont {Y.}~\bibnamefont {Yin}}, \bibinfo {author}
  {\bibfnamefont {A.~N.}\ \bibnamefont {Cleland}},\ and\ \bibinfo {author}
  {\bibfnamefont {J.~M.}\ \bibnamefont {Martinis}},\ }\bibfield  {title}
  {\bibinfo {title} {{Computing prime factors with a Josephson phase qubit
  quantum processor}},\ }\href {https://doi.org/10.1038/nphys2385} {\bibfield
  {journal} {\bibinfo  {journal} {Nature Physics}\ }\textbf {\bibinfo {volume}
  {8}},\ \bibinfo {pages} {719} (\bibinfo {year} {2012})}\BibitemShut {NoStop}%
\bibitem [{\citenamefont {C{\'{o}}rcoles}\ \emph {et~al.}(2015)\citenamefont
  {C{\'{o}}rcoles}, \citenamefont {Magesan}, \citenamefont {Srinivasan},
  \citenamefont {Cross}, \citenamefont {Steffen}, \citenamefont {Gambetta},\
  and\ \citenamefont {Chow}}]{Corcoles2015}%
  \BibitemOpen
  \bibfield  {author} {\bibinfo {author} {\bibfnamefont {A.~D.}\ \bibnamefont
  {C{\'{o}}rcoles}}, \bibinfo {author} {\bibfnamefont {E.}~\bibnamefont
  {Magesan}}, \bibinfo {author} {\bibfnamefont {S.~J.}\ \bibnamefont
  {Srinivasan}}, \bibinfo {author} {\bibfnamefont {A.~W.}\ \bibnamefont
  {Cross}}, \bibinfo {author} {\bibfnamefont {M.}~\bibnamefont {Steffen}},
  \bibinfo {author} {\bibfnamefont {J.~M.}\ \bibnamefont {Gambetta}},\ and\
  \bibinfo {author} {\bibfnamefont {J.~M.}\ \bibnamefont {Chow}},\ }\bibfield
  {title} {\bibinfo {title} {{Demonstration of a quantum error detection code
  using a square lattice of four superconducting qubits}},\ }\href
  {https://doi.org/10.1038/ncomms7979} {\bibfield  {journal} {\bibinfo
  {journal} {Nature Communications}\ }\textbf {\bibinfo {volume} {6}},\
  \bibinfo {pages} {6979} (\bibinfo {year} {2015})}\BibitemShut {NoStop}%
\bibitem [{\citenamefont {Rist{\`{e}}}\ \emph {et~al.}(2015)\citenamefont
  {Rist{\`{e}}}, \citenamefont {Poletto}, \citenamefont {Huang}, \citenamefont
  {Bruno}, \citenamefont {Vesterinen}, \citenamefont {Saira},\ and\
  \citenamefont {DiCarlo}}]{Riste2015}%
  \BibitemOpen
  \bibfield  {author} {\bibinfo {author} {\bibfnamefont {D.}~\bibnamefont
  {Rist{\`{e}}}}, \bibinfo {author} {\bibfnamefont {S.}~\bibnamefont
  {Poletto}}, \bibinfo {author} {\bibfnamefont {M.-Z.}\ \bibnamefont {Huang}},
  \bibinfo {author} {\bibfnamefont {A.}~\bibnamefont {Bruno}}, \bibinfo
  {author} {\bibfnamefont {V.}~\bibnamefont {Vesterinen}}, \bibinfo {author}
  {\bibfnamefont {O.-P.}\ \bibnamefont {Saira}},\ and\ \bibinfo {author}
  {\bibfnamefont {L.}~\bibnamefont {DiCarlo}},\ }\bibfield  {title} {\bibinfo
  {title} {{Detecting bit-flip errors in a logical qubit using stabilizer
  measurements}},\ }\href {https://doi.org/10.1038/ncomms7983} {\bibfield
  {journal} {\bibinfo  {journal} {Nature Communications}\ }\textbf {\bibinfo
  {volume} {6}},\ \bibinfo {pages} {6983} (\bibinfo {year} {2015})}\BibitemShut
  {NoStop}%
\bibitem [{\citenamefont {Barends}\ \emph {et~al.}(2014)\citenamefont
  {Barends}, \citenamefont {Kelly}, \citenamefont {Megrant}, \citenamefont
  {Veitia}, \citenamefont {Sank}, \citenamefont {Jeffrey}, \citenamefont
  {White}, \citenamefont {Mutus}, \citenamefont {Fowler}, \citenamefont
  {Campbell}, \citenamefont {Chen}, \citenamefont {Chen}, \citenamefont
  {Chiaro}, \citenamefont {Dunsworth}, \citenamefont {Neill}, \citenamefont
  {O'Malley}, \citenamefont {Roushan}, \citenamefont {Vainsencher},
  \citenamefont {Wenner}, \citenamefont {Korotkov}, \citenamefont {Cleland},\
  and\ \citenamefont {Martinis}}]{Barends2014}%
  \BibitemOpen
  \bibfield  {author} {\bibinfo {author} {\bibfnamefont {R.}~\bibnamefont
  {Barends}}, \bibinfo {author} {\bibfnamefont {J.}~\bibnamefont {Kelly}},
  \bibinfo {author} {\bibfnamefont {A.}~\bibnamefont {Megrant}}, \bibinfo
  {author} {\bibfnamefont {A.}~\bibnamefont {Veitia}}, \bibinfo {author}
  {\bibfnamefont {D.}~\bibnamefont {Sank}}, \bibinfo {author} {\bibfnamefont
  {E.}~\bibnamefont {Jeffrey}}, \bibinfo {author} {\bibfnamefont {T.~C.}\
  \bibnamefont {White}}, \bibinfo {author} {\bibfnamefont {J.}~\bibnamefont
  {Mutus}}, \bibinfo {author} {\bibfnamefont {A.~G.}\ \bibnamefont {Fowler}},
  \bibinfo {author} {\bibfnamefont {B.}~\bibnamefont {Campbell}}, \bibinfo
  {author} {\bibfnamefont {Y.}~\bibnamefont {Chen}}, \bibinfo {author}
  {\bibfnamefont {Z.}~\bibnamefont {Chen}}, \bibinfo {author} {\bibfnamefont
  {B.}~\bibnamefont {Chiaro}}, \bibinfo {author} {\bibfnamefont
  {A.}~\bibnamefont {Dunsworth}}, \bibinfo {author} {\bibfnamefont
  {C.}~\bibnamefont {Neill}}, \bibinfo {author} {\bibfnamefont
  {P.}~\bibnamefont {O'Malley}}, \bibinfo {author} {\bibfnamefont
  {P.}~\bibnamefont {Roushan}}, \bibinfo {author} {\bibfnamefont
  {A.}~\bibnamefont {Vainsencher}}, \bibinfo {author} {\bibfnamefont
  {J.}~\bibnamefont {Wenner}}, \bibinfo {author} {\bibfnamefont {A.~N.}\
  \bibnamefont {Korotkov}}, \bibinfo {author} {\bibfnamefont {A.~N.}\
  \bibnamefont {Cleland}},\ and\ \bibinfo {author} {\bibfnamefont {J.~M.}\
  \bibnamefont {Martinis}},\ }\bibfield  {title} {\bibinfo {title}
  {{Superconducting quantum circuits at the surface code threshold for fault
  tolerance}},\ }\href {https://doi.org/10.1038/nature13171} {\bibfield
  {journal} {\bibinfo  {journal} {Nature}\ }\textbf {\bibinfo {volume} {508}},\
  \bibinfo {pages} {500} (\bibinfo {year} {2014})}\BibitemShut {NoStop}%
\bibitem [{\citenamefont {Kelly}\ \emph {et~al.}(2015)\citenamefont {Kelly},
  \citenamefont {Barends}, \citenamefont {Fowler}, \citenamefont {Megrant},
  \citenamefont {Jeffrey}, \citenamefont {White}, \citenamefont {Sank},
  \citenamefont {Mutus}, \citenamefont {Campbell}, \citenamefont {Chen},
  \citenamefont {Chen}, \citenamefont {Chiaro}, \citenamefont {Dunsworth},
  \citenamefont {Hoi}, \citenamefont {Neill}, \citenamefont {O'Malley},
  \citenamefont {Quintana}, \citenamefont {Roushan}, \citenamefont
  {Vainsencher}, \citenamefont {Wenner}, \citenamefont {Cleland},\ and\
  \citenamefont {Martinis}}]{Kelly2015}%
  \BibitemOpen
  \bibfield  {author} {\bibinfo {author} {\bibfnamefont {J.}~\bibnamefont
  {Kelly}}, \bibinfo {author} {\bibfnamefont {R.}~\bibnamefont {Barends}},
  \bibinfo {author} {\bibfnamefont {A.~G.}\ \bibnamefont {Fowler}}, \bibinfo
  {author} {\bibfnamefont {A.}~\bibnamefont {Megrant}}, \bibinfo {author}
  {\bibfnamefont {E.}~\bibnamefont {Jeffrey}}, \bibinfo {author} {\bibfnamefont
  {T.~C.}\ \bibnamefont {White}}, \bibinfo {author} {\bibfnamefont
  {D.}~\bibnamefont {Sank}}, \bibinfo {author} {\bibfnamefont {J.}~\bibnamefont
  {Mutus}}, \bibinfo {author} {\bibfnamefont {B.}~\bibnamefont {Campbell}},
  \bibinfo {author} {\bibfnamefont {Y.}~\bibnamefont {Chen}}, \bibinfo {author}
  {\bibfnamefont {Z.}~\bibnamefont {Chen}}, \bibinfo {author} {\bibfnamefont
  {B.}~\bibnamefont {Chiaro}}, \bibinfo {author} {\bibfnamefont
  {A.}~\bibnamefont {Dunsworth}}, \bibinfo {author} {\bibfnamefont {I.-C.}\
  \bibnamefont {Hoi}}, \bibinfo {author} {\bibfnamefont {C.}~\bibnamefont
  {Neill}}, \bibinfo {author} {\bibfnamefont {P.~J.~J.}\ \bibnamefont
  {O'Malley}}, \bibinfo {author} {\bibfnamefont {C.}~\bibnamefont {Quintana}},
  \bibinfo {author} {\bibfnamefont {P.}~\bibnamefont {Roushan}}, \bibinfo
  {author} {\bibfnamefont {A.}~\bibnamefont {Vainsencher}}, \bibinfo {author}
  {\bibfnamefont {J.}~\bibnamefont {Wenner}}, \bibinfo {author} {\bibfnamefont
  {A.~N.}\ \bibnamefont {Cleland}},\ and\ \bibinfo {author} {\bibfnamefont
  {J.~M.}\ \bibnamefont {Martinis}},\ }\bibfield  {title} {\bibinfo {title}
  {{State preservation by repetitive error detection in a superconducting
  quantum circuit}},\ }\href {https://doi.org/10.1038/nature14270} {\bibfield
  {journal} {\bibinfo  {journal} {Nature}\ }\textbf {\bibinfo {volume} {519}},\
  \bibinfo {pages} {66} (\bibinfo {year} {2015})}\BibitemShut {NoStop}%
\bibitem [{\citenamefont {Barends}\ \emph {et~al.}(2013)\citenamefont
  {Barends}, \citenamefont {Kelly}, \citenamefont {Megrant}, \citenamefont
  {Sank}, \citenamefont {Jeffrey}, \citenamefont {Chen}, \citenamefont {Yin},
  \citenamefont {Chiaro}, \citenamefont {Mutus}, \citenamefont {Neill},
  \citenamefont {O'Malley}, \citenamefont {Roushan}, \citenamefont {Wenner},
  \citenamefont {White}, \citenamefont {Cleland},\ and\ \citenamefont
  {Martinis}}]{Barends2013}%
  \BibitemOpen
  \bibfield  {author} {\bibinfo {author} {\bibfnamefont {R.}~\bibnamefont
  {Barends}}, \bibinfo {author} {\bibfnamefont {J.}~\bibnamefont {Kelly}},
  \bibinfo {author} {\bibfnamefont {A.}~\bibnamefont {Megrant}}, \bibinfo
  {author} {\bibfnamefont {D.}~\bibnamefont {Sank}}, \bibinfo {author}
  {\bibfnamefont {E.}~\bibnamefont {Jeffrey}}, \bibinfo {author} {\bibfnamefont
  {Y.}~\bibnamefont {Chen}}, \bibinfo {author} {\bibfnamefont {Y.}~\bibnamefont
  {Yin}}, \bibinfo {author} {\bibfnamefont {B.}~\bibnamefont {Chiaro}},
  \bibinfo {author} {\bibfnamefont {J.}~\bibnamefont {Mutus}}, \bibinfo
  {author} {\bibfnamefont {C.}~\bibnamefont {Neill}}, \bibinfo {author}
  {\bibfnamefont {P.}~\bibnamefont {O'Malley}}, \bibinfo {author}
  {\bibfnamefont {P.}~\bibnamefont {Roushan}}, \bibinfo {author} {\bibfnamefont
  {J.}~\bibnamefont {Wenner}}, \bibinfo {author} {\bibfnamefont {T.~C.}\
  \bibnamefont {White}}, \bibinfo {author} {\bibfnamefont {A.~N.}\ \bibnamefont
  {Cleland}},\ and\ \bibinfo {author} {\bibfnamefont {J.~M.}\ \bibnamefont
  {Martinis}},\ }\bibfield  {title} {\bibinfo {title} {{Coherent Josephson
  Qubit Suitable for Scalable Quantum Integrated Circuits}},\ }\href
  {https://doi.org/10.1103/PhysRevLett.111.080502} {\bibfield  {journal}
  {\bibinfo  {journal} {Physical Review Letters}\ }\textbf {\bibinfo {volume}
  {111}},\ \bibinfo {pages} {080502} (\bibinfo {year} {2013})}\BibitemShut
  {NoStop}%
\bibitem [{\citenamefont {Arute}\ \emph {et~al.}(2019)\citenamefont {Arute},
  \citenamefont {Arya}, \citenamefont {Babbush}, \citenamefont {Bacon},
  \citenamefont {Bardin}, \citenamefont {Barends}, \citenamefont {Biswas},
  \citenamefont {Boixo}, \citenamefont {Brandao}, \citenamefont {Buell},
  \citenamefont {Burkett}, \citenamefont {Chen}, \citenamefont {Chen},
  \citenamefont {Chiaro}, \citenamefont {Collins}, \citenamefont {Courtney},
  \citenamefont {Dunsworth}, \citenamefont {Farhi}, \citenamefont {Foxen},
  \citenamefont {Fowler}, \citenamefont {Gidney}, \citenamefont {Giustina},
  \citenamefont {Graff}, \citenamefont {Guerin}, \citenamefont {Habegger},
  \citenamefont {Harrigan}, \citenamefont {Hartmann}, \citenamefont {Ho},
  \citenamefont {Hoffmann}, \citenamefont {Huang}, \citenamefont {Humble},
  \citenamefont {Isakov}, \citenamefont {Jeffrey}, \citenamefont {Jiang},
  \citenamefont {Kafri}, \citenamefont {Kechedzhi}, \citenamefont {Kelly},
  \citenamefont {Klimov}, \citenamefont {Knysh}, \citenamefont {Korotkov},
  \citenamefont {Kostritsa}, \citenamefont {Landhuis}, \citenamefont
  {Lindmark}, \citenamefont {Lucero}, \citenamefont {Lyakh}, \citenamefont
  {Mandr{\`{a}}}, \citenamefont {McClean}, \citenamefont {McEwen},
  \citenamefont {Megrant}, \citenamefont {Mi}, \citenamefont {Michielsen},
  \citenamefont {Mohseni}, \citenamefont {Mutus}, \citenamefont {Naaman},
  \citenamefont {Neeley}, \citenamefont {Neill}, \citenamefont {Niu},
  \citenamefont {Ostby}, \citenamefont {Petukhov}, \citenamefont {Platt},
  \citenamefont {Quintana}, \citenamefont {Rieffel}, \citenamefont {Roushan},
  \citenamefont {Rubin}, \citenamefont {Sank}, \citenamefont {Satzinger},
  \citenamefont {Smelyanskiy}, \citenamefont {Sung}, \citenamefont
  {Trevithick}, \citenamefont {Vainsencher}, \citenamefont {Villalonga},
  \citenamefont {White}, \citenamefont {Yao}, \citenamefont {Yeh},
  \citenamefont {Zalcman}, \citenamefont {Neven},\ and\ \citenamefont
  {Martinis}}]{Arute2019}%
  \BibitemOpen
  \bibfield  {author} {\bibinfo {author} {\bibfnamefont {F.}~\bibnamefont
  {Arute}}, \bibinfo {author} {\bibfnamefont {K.}~\bibnamefont {Arya}},
  \bibinfo {author} {\bibfnamefont {R.}~\bibnamefont {Babbush}}, \bibinfo
  {author} {\bibfnamefont {D.}~\bibnamefont {Bacon}}, \bibinfo {author}
  {\bibfnamefont {J.~C.}\ \bibnamefont {Bardin}}, \bibinfo {author}
  {\bibfnamefont {R.}~\bibnamefont {Barends}}, \bibinfo {author} {\bibfnamefont
  {R.}~\bibnamefont {Biswas}}, \bibinfo {author} {\bibfnamefont
  {S.}~\bibnamefont {Boixo}}, \bibinfo {author} {\bibfnamefont {F.~G. S.~L.}\
  \bibnamefont {Brandao}}, \bibinfo {author} {\bibfnamefont {D.~A.}\
  \bibnamefont {Buell}}, \bibinfo {author} {\bibfnamefont {B.}~\bibnamefont
  {Burkett}}, \bibinfo {author} {\bibfnamefont {Y.}~\bibnamefont {Chen}},
  \bibinfo {author} {\bibfnamefont {Z.}~\bibnamefont {Chen}}, \bibinfo {author}
  {\bibfnamefont {B.}~\bibnamefont {Chiaro}}, \bibinfo {author} {\bibfnamefont
  {R.}~\bibnamefont {Collins}}, \bibinfo {author} {\bibfnamefont
  {W.}~\bibnamefont {Courtney}}, \bibinfo {author} {\bibfnamefont
  {A.}~\bibnamefont {Dunsworth}}, \bibinfo {author} {\bibfnamefont
  {E.}~\bibnamefont {Farhi}}, \bibinfo {author} {\bibfnamefont
  {B.}~\bibnamefont {Foxen}}, \bibinfo {author} {\bibfnamefont
  {A.}~\bibnamefont {Fowler}}, \bibinfo {author} {\bibfnamefont
  {C.}~\bibnamefont {Gidney}}, \bibinfo {author} {\bibfnamefont
  {M.}~\bibnamefont {Giustina}}, \bibinfo {author} {\bibfnamefont
  {R.}~\bibnamefont {Graff}}, \bibinfo {author} {\bibfnamefont
  {K.}~\bibnamefont {Guerin}}, \bibinfo {author} {\bibfnamefont
  {S.}~\bibnamefont {Habegger}}, \bibinfo {author} {\bibfnamefont {M.~P.}\
  \bibnamefont {Harrigan}}, \bibinfo {author} {\bibfnamefont {M.~J.}\
  \bibnamefont {Hartmann}}, \bibinfo {author} {\bibfnamefont {A.}~\bibnamefont
  {Ho}}, \bibinfo {author} {\bibfnamefont {M.}~\bibnamefont {Hoffmann}},
  \bibinfo {author} {\bibfnamefont {T.}~\bibnamefont {Huang}}, \bibinfo
  {author} {\bibfnamefont {T.~S.}\ \bibnamefont {Humble}}, \bibinfo {author}
  {\bibfnamefont {S.~V.}\ \bibnamefont {Isakov}}, \bibinfo {author}
  {\bibfnamefont {E.}~\bibnamefont {Jeffrey}}, \bibinfo {author} {\bibfnamefont
  {Z.}~\bibnamefont {Jiang}}, \bibinfo {author} {\bibfnamefont
  {D.}~\bibnamefont {Kafri}}, \bibinfo {author} {\bibfnamefont
  {K.}~\bibnamefont {Kechedzhi}}, \bibinfo {author} {\bibfnamefont
  {J.}~\bibnamefont {Kelly}}, \bibinfo {author} {\bibfnamefont {P.~V.}\
  \bibnamefont {Klimov}}, \bibinfo {author} {\bibfnamefont {S.}~\bibnamefont
  {Knysh}}, \bibinfo {author} {\bibfnamefont {A.~N.}\ \bibnamefont {Korotkov}},
  \bibinfo {author} {\bibfnamefont {F.}~\bibnamefont {Kostritsa}}, \bibinfo
  {author} {\bibfnamefont {D.}~\bibnamefont {Landhuis}}, \bibinfo {author}
  {\bibfnamefont {M.}~\bibnamefont {Lindmark}}, \bibinfo {author}
  {\bibfnamefont {E.}~\bibnamefont {Lucero}}, \bibinfo {author} {\bibfnamefont
  {D.}~\bibnamefont {Lyakh}}, \bibinfo {author} {\bibfnamefont
  {S.}~\bibnamefont {Mandr{\`{a}}}}, \bibinfo {author} {\bibfnamefont {J.~R.}\
  \bibnamefont {McClean}}, \bibinfo {author} {\bibfnamefont {M.}~\bibnamefont
  {McEwen}}, \bibinfo {author} {\bibfnamefont {A.}~\bibnamefont {Megrant}},
  \bibinfo {author} {\bibfnamefont {X.}~\bibnamefont {Mi}}, \bibinfo {author}
  {\bibfnamefont {K.}~\bibnamefont {Michielsen}}, \bibinfo {author}
  {\bibfnamefont {M.}~\bibnamefont {Mohseni}}, \bibinfo {author} {\bibfnamefont
  {J.}~\bibnamefont {Mutus}}, \bibinfo {author} {\bibfnamefont
  {O.}~\bibnamefont {Naaman}}, \bibinfo {author} {\bibfnamefont
  {M.}~\bibnamefont {Neeley}}, \bibinfo {author} {\bibfnamefont
  {C.}~\bibnamefont {Neill}}, \bibinfo {author} {\bibfnamefont {M.~Y.}\
  \bibnamefont {Niu}}, \bibinfo {author} {\bibfnamefont {E.}~\bibnamefont
  {Ostby}}, \bibinfo {author} {\bibfnamefont {A.}~\bibnamefont {Petukhov}},
  \bibinfo {author} {\bibfnamefont {J.~C.}\ \bibnamefont {Platt}}, \bibinfo
  {author} {\bibfnamefont {C.}~\bibnamefont {Quintana}}, \bibinfo {author}
  {\bibfnamefont {E.~G.}\ \bibnamefont {Rieffel}}, \bibinfo {author}
  {\bibfnamefont {P.}~\bibnamefont {Roushan}}, \bibinfo {author} {\bibfnamefont
  {N.~C.}\ \bibnamefont {Rubin}}, \bibinfo {author} {\bibfnamefont
  {D.}~\bibnamefont {Sank}}, \bibinfo {author} {\bibfnamefont {K.~J.}\
  \bibnamefont {Satzinger}}, \bibinfo {author} {\bibfnamefont {V.}~\bibnamefont
  {Smelyanskiy}}, \bibinfo {author} {\bibfnamefont {K.~J.}\ \bibnamefont
  {Sung}}, \bibinfo {author} {\bibfnamefont {M.~D.}\ \bibnamefont
  {Trevithick}}, \bibinfo {author} {\bibfnamefont {A.}~\bibnamefont
  {Vainsencher}}, \bibinfo {author} {\bibfnamefont {B.}~\bibnamefont
  {Villalonga}}, \bibinfo {author} {\bibfnamefont {T.}~\bibnamefont {White}},
  \bibinfo {author} {\bibfnamefont {Z.~J.}\ \bibnamefont {Yao}}, \bibinfo
  {author} {\bibfnamefont {P.}~\bibnamefont {Yeh}}, \bibinfo {author}
  {\bibfnamefont {A.}~\bibnamefont {Zalcman}}, \bibinfo {author} {\bibfnamefont
  {H.}~\bibnamefont {Neven}},\ and\ \bibinfo {author} {\bibfnamefont {J.~M.}\
  \bibnamefont {Martinis}},\ }\bibfield  {title} {\bibinfo {title} {{Quantum
  supremacy using a programmable superconducting processor}},\ }\href
  {https://doi.org/10.1038/s41586-019-1666-5} {\bibfield  {journal} {\bibinfo
  {journal} {Nature}\ }\textbf {\bibinfo {volume} {574}},\ \bibinfo {pages}
  {505} (\bibinfo {year} {2019})}\BibitemShut {NoStop}%
\bibitem [{\citenamefont {Preskill}(2018)}]{Preskill2018}%
  \BibitemOpen
  \bibfield  {author} {\bibinfo {author} {\bibfnamefont {J.}~\bibnamefont
  {Preskill}},\ }\bibfield  {title} {\bibinfo {title} {{Quantum Computing in
  the NISQ era and beyond}},\ }\href {https://doi.org/10.22331/q-2018-08-06-79}
  {\bibfield  {journal} {\bibinfo  {journal} {Quantum}\ }\textbf {\bibinfo
  {volume} {2}},\ \bibinfo {pages} {79} (\bibinfo {year} {2018})}\BibitemShut
  {NoStop}%
\bibitem [{\citenamefont {DiCarlo}\ \emph {et~al.}(2009)\citenamefont
  {DiCarlo}, \citenamefont {Chow}, \citenamefont {Gambetta}, \citenamefont
  {Bishop}, \citenamefont {Johnson}, \citenamefont {Schuster}, \citenamefont
  {Majer}, \citenamefont {Blais}, \citenamefont {Frunzio}, \citenamefont
  {Girvin},\ and\ \citenamefont {Schoelkopf}}]{DiCarlo2009}%
  \BibitemOpen
  \bibfield  {author} {\bibinfo {author} {\bibfnamefont {L.}~\bibnamefont
  {DiCarlo}}, \bibinfo {author} {\bibfnamefont {J.~M.}\ \bibnamefont {Chow}},
  \bibinfo {author} {\bibfnamefont {J.~M.}\ \bibnamefont {Gambetta}}, \bibinfo
  {author} {\bibfnamefont {L.~S.}\ \bibnamefont {Bishop}}, \bibinfo {author}
  {\bibfnamefont {B.~R.}\ \bibnamefont {Johnson}}, \bibinfo {author}
  {\bibfnamefont {D.~I.}\ \bibnamefont {Schuster}}, \bibinfo {author}
  {\bibfnamefont {J.}~\bibnamefont {Majer}}, \bibinfo {author} {\bibfnamefont
  {A.}~\bibnamefont {Blais}}, \bibinfo {author} {\bibfnamefont
  {L.}~\bibnamefont {Frunzio}}, \bibinfo {author} {\bibfnamefont {S.~M.}\
  \bibnamefont {Girvin}},\ and\ \bibinfo {author} {\bibfnamefont {R.~J.}\
  \bibnamefont {Schoelkopf}},\ }\bibfield  {title} {\bibinfo {title}
  {{Demonstration of two-qubit algorithms with a superconducting quantum
  processor}},\ }\href {https://doi.org/10.1038/nature08121} {\bibfield
  {journal} {\bibinfo  {journal} {Nature}\ }\textbf {\bibinfo {volume} {460}},\
  \bibinfo {pages} {240} (\bibinfo {year} {2009})}\BibitemShut {NoStop}%
\bibitem [{\citenamefont {Yamamoto}\ \emph {et~al.}(2010)\citenamefont
  {Yamamoto}, \citenamefont {Neeley}, \citenamefont {Lucero}, \citenamefont
  {Bialczak}, \citenamefont {Kelly}, \citenamefont {Lenander}, \citenamefont
  {Mariantoni}, \citenamefont {O'Connell}, \citenamefont {Sank}, \citenamefont
  {Wang}, \citenamefont {Weides}, \citenamefont {Wenner}, \citenamefont {Yin},
  \citenamefont {Cleland},\ and\ \citenamefont {Martinis}}]{Yamamoto2010}%
  \BibitemOpen
  \bibfield  {author} {\bibinfo {author} {\bibfnamefont {T.}~\bibnamefont
  {Yamamoto}}, \bibinfo {author} {\bibfnamefont {M.}~\bibnamefont {Neeley}},
  \bibinfo {author} {\bibfnamefont {E.}~\bibnamefont {Lucero}}, \bibinfo
  {author} {\bibfnamefont {R.~C.}\ \bibnamefont {Bialczak}}, \bibinfo {author}
  {\bibfnamefont {J.}~\bibnamefont {Kelly}}, \bibinfo {author} {\bibfnamefont
  {M.}~\bibnamefont {Lenander}}, \bibinfo {author} {\bibfnamefont
  {M.}~\bibnamefont {Mariantoni}}, \bibinfo {author} {\bibfnamefont {A.~D.}\
  \bibnamefont {O'Connell}}, \bibinfo {author} {\bibfnamefont {D.}~\bibnamefont
  {Sank}}, \bibinfo {author} {\bibfnamefont {H.}~\bibnamefont {Wang}}, \bibinfo
  {author} {\bibfnamefont {M.}~\bibnamefont {Weides}}, \bibinfo {author}
  {\bibfnamefont {J.}~\bibnamefont {Wenner}}, \bibinfo {author} {\bibfnamefont
  {Y.}~\bibnamefont {Yin}}, \bibinfo {author} {\bibfnamefont {A.~N.}\
  \bibnamefont {Cleland}},\ and\ \bibinfo {author} {\bibfnamefont {J.~M.}\
  \bibnamefont {Martinis}},\ }\bibfield  {title} {\bibinfo {title} {{Quantum
  process tomography of two-qubit controlled-Z and controlled-NOT gates using
  superconducting phase qubits}},\ }\href
  {https://doi.org/10.1103/PhysRevB.82.184515} {\bibfield  {journal} {\bibinfo
  {journal} {Physical Review B}\ }\textbf {\bibinfo {volume} {82}},\ \bibinfo
  {pages} {184515} (\bibinfo {year} {2010})}\BibitemShut {NoStop}%
\bibitem [{\citenamefont {M{\"{u}}ller}\ \emph {et~al.}(2019)\citenamefont
  {M{\"{u}}ller}, \citenamefont {Cole},\ and\ \citenamefont
  {Lisenfeld}}]{Muller2019}%
  \BibitemOpen
  \bibfield  {author} {\bibinfo {author} {\bibfnamefont {C.}~\bibnamefont
  {M{\"{u}}ller}}, \bibinfo {author} {\bibfnamefont {J.~H.}\ \bibnamefont
  {Cole}},\ and\ \bibinfo {author} {\bibfnamefont {J.}~\bibnamefont
  {Lisenfeld}},\ }\bibfield  {title} {\bibinfo {title} {{Towards understanding
  two-level-systems in amorphous solids: insights from quantum circuits}},\
  }\href {https://doi.org/10.1088/1361-6633/ab3a7e} {\bibfield  {journal}
  {\bibinfo  {journal} {Reports on Progress in Physics}\ }\textbf {\bibinfo
  {volume} {82}},\ \bibinfo {pages} {124501} (\bibinfo {year}
  {2019})}\BibitemShut {NoStop}%
\bibitem [{\citenamefont {Earnest}\ \emph {et~al.}(2018)\citenamefont
  {Earnest}, \citenamefont {B{\'{e}}janin}, \citenamefont {McConkey},
  \citenamefont {Peters}, \citenamefont {Korinek}, \citenamefont {Yuan},\ and\
  \citenamefont {Mariantoni}}]{Earnest2018}%
  \BibitemOpen
  \bibfield  {author} {\bibinfo {author} {\bibfnamefont {C.~T.}\ \bibnamefont
  {Earnest}}, \bibinfo {author} {\bibfnamefont {J.~H.}\ \bibnamefont
  {B{\'{e}}janin}}, \bibinfo {author} {\bibfnamefont {T.~G.}\ \bibnamefont
  {McConkey}}, \bibinfo {author} {\bibfnamefont {E.~A.}\ \bibnamefont
  {Peters}}, \bibinfo {author} {\bibfnamefont {A.}~\bibnamefont {Korinek}},
  \bibinfo {author} {\bibfnamefont {H.}~\bibnamefont {Yuan}},\ and\ \bibinfo
  {author} {\bibfnamefont {M.}~\bibnamefont {Mariantoni}},\ }\bibfield  {title}
  {\bibinfo {title} {{Substrate surface engineering for high-quality
  silicon/aluminum superconducting resonators}},\ }\href
  {https://doi.org/10.1088/1361-6668/aae548} {\bibfield  {journal} {\bibinfo
  {journal} {Superconductor Science and Technology}\ }\textbf {\bibinfo
  {volume} {31}},\ \bibinfo {pages} {125013} (\bibinfo {year}
  {2018})}\BibitemShut {NoStop}%
\bibitem [{\citenamefont {Moeed}\ \emph {et~al.}(2019)\citenamefont {Moeed},
  \citenamefont {Earnest}, \citenamefont {B{\'{e}}janin}, \citenamefont
  {Sharafeldin},\ and\ \citenamefont {Mariantoni}}]{Moeed2019}%
  \BibitemOpen
  \bibfield  {author} {\bibinfo {author} {\bibfnamefont {M.}~\bibnamefont
  {Moeed}}, \bibinfo {author} {\bibfnamefont {C.~T.}\ \bibnamefont {Earnest}},
  \bibinfo {author} {\bibfnamefont {J.}~\bibnamefont {B{\'{e}}janin}}, \bibinfo
  {author} {\bibfnamefont {A.}~\bibnamefont {Sharafeldin}},\ and\ \bibinfo
  {author} {\bibfnamefont {M.}~\bibnamefont {Mariantoni}},\ }\bibfield  {title}
  {\bibinfo {title} {{Improving the Time Stability of Superconducting Planar
  Resonators}},\ }\href {https://doi.org/10.1557/adv.2019.262} {\bibfield
  {journal} {\bibinfo  {journal} {MRS Advances}\ ,\ \bibinfo {pages} {1}}
  (\bibinfo {year} {2019})}\BibitemShut {NoStop}%
\bibitem [{\citenamefont {McConkey}\ \emph {et~al.}(2018)\citenamefont
  {McConkey}, \citenamefont {B{\'{e}}janin}, \citenamefont {Earnest},
  \citenamefont {McRae}, \citenamefont {Pagel}, \citenamefont {Rinehart},\ and\
  \citenamefont {Mariantoni}}]{McConkey2018}%
  \BibitemOpen
  \bibfield  {author} {\bibinfo {author} {\bibfnamefont {T.~G.}\ \bibnamefont
  {McConkey}}, \bibinfo {author} {\bibfnamefont {J.~H.}\ \bibnamefont
  {B{\'{e}}janin}}, \bibinfo {author} {\bibfnamefont {C.~T.}\ \bibnamefont
  {Earnest}}, \bibinfo {author} {\bibfnamefont {C.~R.~H.}\ \bibnamefont
  {McRae}}, \bibinfo {author} {\bibfnamefont {Z.}~\bibnamefont {Pagel}},
  \bibinfo {author} {\bibfnamefont {J.~R.}\ \bibnamefont {Rinehart}},\ and\
  \bibinfo {author} {\bibfnamefont {M.}~\bibnamefont {Mariantoni}},\ }\bibfield
   {title} {\bibinfo {title} {{Mitigating leakage errors due to cavity modes in
  a superconducting quantum computer}},\ }\href
  {https://doi.org/10.1088/2058-9565/aabd41} {\bibfield  {journal} {\bibinfo
  {journal} {Quantum Science and Technology}\ }\textbf {\bibinfo {volume}
  {3}},\ \bibinfo {pages} {034004} (\bibinfo {year} {2018})}\BibitemShut
  {NoStop}%
\bibitem [{\citenamefont {Klimov}\ \emph {et~al.}(2018)\citenamefont {Klimov},
  \citenamefont {Kelly}, \citenamefont {Chen}, \citenamefont {Neeley},
  \citenamefont {Megrant}, \citenamefont {Burkett}, \citenamefont {Barends},
  \citenamefont {Arya}, \citenamefont {Chiaro}, \citenamefont {Chen},
  \citenamefont {Dunsworth}, \citenamefont {Fowler}, \citenamefont {Foxen},
  \citenamefont {Gidney}, \citenamefont {Giustina}, \citenamefont {Graff},
  \citenamefont {Huang}, \citenamefont {Jeffrey}, \citenamefont {Lucero},
  \citenamefont {Mutus}, \citenamefont {Naaman}, \citenamefont {Neill},
  \citenamefont {Quintana}, \citenamefont {Roushan}, \citenamefont {Sank},
  \citenamefont {Vainsencher}, \citenamefont {Wenner}, \citenamefont {White},
  \citenamefont {Boixo}, \citenamefont {Babbush}, \citenamefont {Smelyanskiy},
  \citenamefont {Neven},\ and\ \citenamefont {Martinis}}]{Klimov2018}%
  \BibitemOpen
  \bibfield  {author} {\bibinfo {author} {\bibfnamefont {P.~V.}\ \bibnamefont
  {Klimov}}, \bibinfo {author} {\bibfnamefont {J.}~\bibnamefont {Kelly}},
  \bibinfo {author} {\bibfnamefont {Z.}~\bibnamefont {Chen}}, \bibinfo {author}
  {\bibfnamefont {M.}~\bibnamefont {Neeley}}, \bibinfo {author} {\bibfnamefont
  {A.}~\bibnamefont {Megrant}}, \bibinfo {author} {\bibfnamefont
  {B.}~\bibnamefont {Burkett}}, \bibinfo {author} {\bibfnamefont
  {R.}~\bibnamefont {Barends}}, \bibinfo {author} {\bibfnamefont
  {K.}~\bibnamefont {Arya}}, \bibinfo {author} {\bibfnamefont {B.}~\bibnamefont
  {Chiaro}}, \bibinfo {author} {\bibfnamefont {Y.}~\bibnamefont {Chen}},
  \bibinfo {author} {\bibfnamefont {A.}~\bibnamefont {Dunsworth}}, \bibinfo
  {author} {\bibfnamefont {A.~G.}\ \bibnamefont {Fowler}}, \bibinfo {author}
  {\bibfnamefont {B.}~\bibnamefont {Foxen}}, \bibinfo {author} {\bibfnamefont
  {C.}~\bibnamefont {Gidney}}, \bibinfo {author} {\bibfnamefont
  {M.}~\bibnamefont {Giustina}}, \bibinfo {author} {\bibfnamefont
  {R.}~\bibnamefont {Graff}}, \bibinfo {author} {\bibfnamefont
  {T.}~\bibnamefont {Huang}}, \bibinfo {author} {\bibfnamefont
  {E.}~\bibnamefont {Jeffrey}}, \bibinfo {author} {\bibfnamefont
  {E.}~\bibnamefont {Lucero}}, \bibinfo {author} {\bibfnamefont
  {J.}~\bibnamefont {Mutus}}, \bibinfo {author} {\bibfnamefont
  {O.}~\bibnamefont {Naaman}}, \bibinfo {author} {\bibfnamefont
  {C.}~\bibnamefont {Neill}}, \bibinfo {author} {\bibfnamefont
  {C.}~\bibnamefont {Quintana}}, \bibinfo {author} {\bibfnamefont
  {P.}~\bibnamefont {Roushan}}, \bibinfo {author} {\bibfnamefont
  {D.}~\bibnamefont {Sank}}, \bibinfo {author} {\bibfnamefont {A.}~\bibnamefont
  {Vainsencher}}, \bibinfo {author} {\bibfnamefont {J.}~\bibnamefont {Wenner}},
  \bibinfo {author} {\bibfnamefont {T.~C.}\ \bibnamefont {White}}, \bibinfo
  {author} {\bibfnamefont {S.}~\bibnamefont {Boixo}}, \bibinfo {author}
  {\bibfnamefont {R.}~\bibnamefont {Babbush}}, \bibinfo {author} {\bibfnamefont
  {V.~N.}\ \bibnamefont {Smelyanskiy}}, \bibinfo {author} {\bibfnamefont
  {H.}~\bibnamefont {Neven}},\ and\ \bibinfo {author} {\bibfnamefont {J.~M.}\
  \bibnamefont {Martinis}},\ }\bibfield  {title} {\bibinfo {title}
  {{Fluctuations of Energy-Relaxation Times in Superconducting Qubits}},\
  }\href {https://doi.org/10.1103/PhysRevLett.121.090502} {\bibfield  {journal}
  {\bibinfo  {journal} {Physical Review Letters}\ }\textbf {\bibinfo {volume}
  {121}},\ \bibinfo {pages} {90502} (\bibinfo {year} {2018})}\BibitemShut
  {NoStop}%
\bibitem [{\citenamefont {Mariantoni}\ \emph
  {et~al.}(2011{\natexlab{b}})\citenamefont {Mariantoni}, \citenamefont {Wang},
  \citenamefont {Bialczak}, \citenamefont {Lenander}, \citenamefont {Lucero},
  \citenamefont {Neeley}, \citenamefont {O'Connell}, \citenamefont {Sank},
  \citenamefont {Weides}, \citenamefont {Wenner}, \citenamefont {Yamamoto},
  \citenamefont {Yin}, \citenamefont {Zhao}, \citenamefont {Martinis},\ and\
  \citenamefont {Cleland}}]{Mariantoni2011a}%
  \BibitemOpen
  \bibfield  {author} {\bibinfo {author} {\bibfnamefont {M.}~\bibnamefont
  {Mariantoni}}, \bibinfo {author} {\bibfnamefont {H.}~\bibnamefont {Wang}},
  \bibinfo {author} {\bibfnamefont {R.~C.}\ \bibnamefont {Bialczak}}, \bibinfo
  {author} {\bibfnamefont {M.}~\bibnamefont {Lenander}}, \bibinfo {author}
  {\bibfnamefont {E.}~\bibnamefont {Lucero}}, \bibinfo {author} {\bibfnamefont
  {M.}~\bibnamefont {Neeley}}, \bibinfo {author} {\bibfnamefont {A.~D.}\
  \bibnamefont {O'Connell}}, \bibinfo {author} {\bibfnamefont {D.}~\bibnamefont
  {Sank}}, \bibinfo {author} {\bibfnamefont {M.}~\bibnamefont {Weides}},
  \bibinfo {author} {\bibfnamefont {J.}~\bibnamefont {Wenner}}, \bibinfo
  {author} {\bibfnamefont {T.}~\bibnamefont {Yamamoto}}, \bibinfo {author}
  {\bibfnamefont {Y.}~\bibnamefont {Yin}}, \bibinfo {author} {\bibfnamefont
  {J.}~\bibnamefont {Zhao}}, \bibinfo {author} {\bibfnamefont {J.~M.}\
  \bibnamefont {Martinis}},\ and\ \bibinfo {author} {\bibfnamefont {A.~N.}\
  \bibnamefont {Cleland}},\ }\bibfield  {title} {\bibinfo {title} {{Photon
  shell game in three-resonator circuit quantum electrodynamics}},\ }\href
  {https://doi.org/10.1038/nphys1885} {\bibfield  {journal} {\bibinfo
  {journal} {Nature Physics}\ }\textbf {\bibinfo {volume} {7}},\ \bibinfo
  {pages} {287} (\bibinfo {year} {2011}{\natexlab{b}})}\BibitemShut {NoStop}%
\bibitem [{\citenamefont {Sanders}(2016)}]{Sanders2016}%
  \BibitemOpen
  \bibfield  {author} {\bibinfo {author} {\bibfnamefont {Y.~R.}\ \bibnamefont
  {Sanders}},\ }\emph {\bibinfo {title} {{Characterizing Errors in Quantum
  Information Processors}}},\ \href {http://hdl.handle.net/10012/10467} {Ph.D.
  thesis},\ \bibinfo  {school} {University of Waterloo} (\bibinfo {year}
  {2016})\BibitemShut {NoStop}%
\bibitem [{\citenamefont {Stenberg}\ \emph {et~al.}(2014)\citenamefont
  {Stenberg}, \citenamefont {Sanders},\ and\ \citenamefont
  {Wilhelm}}]{Stenberg2014}%
  \BibitemOpen
  \bibfield  {author} {\bibinfo {author} {\bibfnamefont {M.~P.}\ \bibnamefont
  {Stenberg}}, \bibinfo {author} {\bibfnamefont {Y.~R.}\ \bibnamefont
  {Sanders}},\ and\ \bibinfo {author} {\bibfnamefont {F.~K.}\ \bibnamefont
  {Wilhelm}},\ }\bibfield  {title} {\bibinfo {title} {{Efficient Estimation of
  Resonant Coupling between Quantum Systems}},\ }\href
  {https://doi.org/10.1103/PhysRevLett.113.210404} {\bibfield  {journal}
  {\bibinfo  {journal} {Physical Review Letters}\ }\textbf {\bibinfo {volume}
  {113}},\ \bibinfo {pages} {210404} (\bibinfo {year} {2014})}\BibitemShut
  {NoStop}%
\bibitem [{\citenamefont {Granade}\ \emph {et~al.}(2012)\citenamefont
  {Granade}, \citenamefont {Ferrie}, \citenamefont {Wiebe},\ and\ \citenamefont
  {Cory}}]{Granade2012}%
  \BibitemOpen
  \bibfield  {author} {\bibinfo {author} {\bibfnamefont {C.~E.}\ \bibnamefont
  {Granade}}, \bibinfo {author} {\bibfnamefont {C.}~\bibnamefont {Ferrie}},
  \bibinfo {author} {\bibfnamefont {N.}~\bibnamefont {Wiebe}},\ and\ \bibinfo
  {author} {\bibfnamefont {D.~G.}\ \bibnamefont {Cory}},\ }\bibfield  {title}
  {\bibinfo {title} {{Robust online Hamiltonian learning}},\ }\href
  {https://doi.org/10.1088/1367-2630/14/10/103013} {\bibfield  {journal}
  {\bibinfo  {journal} {New Journal of Physics}\ }\textbf {\bibinfo {volume}
  {14}},\ \bibinfo {pages} {103013} (\bibinfo {year} {2012})}\BibitemShut
  {NoStop}%
\bibitem [{\citenamefont {Koch}\ \emph {et~al.}(2007)\citenamefont {Koch},
  \citenamefont {Yu}, \citenamefont {Gambetta}, \citenamefont {Houck},
  \citenamefont {Schuster}, \citenamefont {Majer}, \citenamefont {Blais},
  \citenamefont {Devoret}, \citenamefont {Girvin},\ and\ \citenamefont
  {Schoelkopf}}]{Koch2007}%
  \BibitemOpen
  \bibfield  {author} {\bibinfo {author} {\bibfnamefont {J.}~\bibnamefont
  {Koch}}, \bibinfo {author} {\bibfnamefont {T.}~\bibnamefont {Yu}}, \bibinfo
  {author} {\bibfnamefont {J.~M.}\ \bibnamefont {Gambetta}}, \bibinfo {author}
  {\bibfnamefont {A.~A.}\ \bibnamefont {Houck}}, \bibinfo {author}
  {\bibfnamefont {D.~I.}\ \bibnamefont {Schuster}}, \bibinfo {author}
  {\bibfnamefont {J.}~\bibnamefont {Majer}}, \bibinfo {author} {\bibfnamefont
  {A.}~\bibnamefont {Blais}}, \bibinfo {author} {\bibfnamefont {M.~H.}\
  \bibnamefont {Devoret}}, \bibinfo {author} {\bibfnamefont {S.~M.}\
  \bibnamefont {Girvin}},\ and\ \bibinfo {author} {\bibfnamefont {R.~J.}\
  \bibnamefont {Schoelkopf}},\ }\bibfield  {title} {\bibinfo {title}
  {{Charge-insensitive qubit design derived from the Cooper pair box}},\ }\href
  {https://doi.org/10.1103/PhysRevA.76.042319} {\bibfield  {journal} {\bibinfo
  {journal} {Physical Review A}\ }\textbf {\bibinfo {volume} {76}},\ \bibinfo
  {pages} {042319} (\bibinfo {year} {2007})}\BibitemShut {NoStop}%
\bibitem [{\citenamefont {Haack}\ \emph {et~al.}(2010)\citenamefont {Haack},
  \citenamefont {Helmer}, \citenamefont {Mariantoni}, \citenamefont
  {Marquardt},\ and\ \citenamefont {Solano}}]{Haack2010}%
  \BibitemOpen
  \bibfield  {author} {\bibinfo {author} {\bibfnamefont {G.}~\bibnamefont
  {Haack}}, \bibinfo {author} {\bibfnamefont {F.}~\bibnamefont {Helmer}},
  \bibinfo {author} {\bibfnamefont {M.}~\bibnamefont {Mariantoni}}, \bibinfo
  {author} {\bibfnamefont {F.}~\bibnamefont {Marquardt}},\ and\ \bibinfo
  {author} {\bibfnamefont {E.}~\bibnamefont {Solano}},\ }\bibfield  {title}
  {\bibinfo {title} {{Resonant quantum gates in circuit quantum
  electrodynamics}},\ }\href {https://doi.org/10.1103/PhysRevB.82.024514}
  {\bibfield  {journal} {\bibinfo  {journal} {Physical Review B}\ }\textbf
  {\bibinfo {volume} {82}},\ \bibinfo {pages} {024514} (\bibinfo {year}
  {2010})}\BibitemShut {NoStop}%
\bibitem [{\citenamefont {Krantz}\ \emph {et~al.}(2019)\citenamefont {Krantz},
  \citenamefont {Kjaergaard}, \citenamefont {Yan}, \citenamefont {Orlando},
  \citenamefont {Gustavsson},\ and\ \citenamefont {Oliver}}]{Krantz2019}%
  \BibitemOpen
  \bibfield  {author} {\bibinfo {author} {\bibfnamefont {P.}~\bibnamefont
  {Krantz}}, \bibinfo {author} {\bibfnamefont {M.}~\bibnamefont {Kjaergaard}},
  \bibinfo {author} {\bibfnamefont {F.}~\bibnamefont {Yan}}, \bibinfo {author}
  {\bibfnamefont {T.~P.}\ \bibnamefont {Orlando}}, \bibinfo {author}
  {\bibfnamefont {S.}~\bibnamefont {Gustavsson}},\ and\ \bibinfo {author}
  {\bibfnamefont {W.~D.}\ \bibnamefont {Oliver}},\ }\bibfield  {title}
  {\bibinfo {title} {{A quantum engineer's guide to superconducting qubits}},\
  }\href {https://doi.org/10.1063/1.5089550} {\bibfield  {journal} {\bibinfo
  {journal} {Applied Physics Reviews}\ }\textbf {\bibinfo {volume} {6}},\
  \bibinfo {pages} {021318} (\bibinfo {year} {2019})}\BibitemShut {NoStop}%
\bibitem [{\citenamefont {Schl{\"{o}}r}\ \emph {et~al.}(2019)\citenamefont
  {Schl{\"{o}}r}, \citenamefont {Lisenfeld}, \citenamefont {M{\"{u}}ller},
  \citenamefont {Bilmes}, \citenamefont {Schneider}, \citenamefont {Pappas},
  \citenamefont {Ustinov},\ and\ \citenamefont {Weides}}]{Schlor2019}%
  \BibitemOpen
  \bibfield  {author} {\bibinfo {author} {\bibfnamefont {S.}~\bibnamefont
  {Schl{\"{o}}r}}, \bibinfo {author} {\bibfnamefont {J.}~\bibnamefont
  {Lisenfeld}}, \bibinfo {author} {\bibfnamefont {C.}~\bibnamefont
  {M{\"{u}}ller}}, \bibinfo {author} {\bibfnamefont {A.}~\bibnamefont
  {Bilmes}}, \bibinfo {author} {\bibfnamefont {A.}~\bibnamefont {Schneider}},
  \bibinfo {author} {\bibfnamefont {D.~P.}\ \bibnamefont {Pappas}}, \bibinfo
  {author} {\bibfnamefont {A.~V.}\ \bibnamefont {Ustinov}},\ and\ \bibinfo
  {author} {\bibfnamefont {M.}~\bibnamefont {Weides}},\ }\bibfield  {title}
  {\bibinfo {title} {{Correlating Decoherence in Transmon Qubits: Low Frequency
  Noise by Single Fluctuators}},\ }\href
  {https://doi.org/10.1103/PhysRevLett.123.190502} {\bibfield  {journal}
  {\bibinfo  {journal} {Physical Review Letters}\ }\textbf {\bibinfo {volume}
  {123}},\ \bibinfo {pages} {190502} (\bibinfo {year} {2019})}\BibitemShut
  {NoStop}%
\bibitem [{\citenamefont {Burnett}\ \emph {et~al.}(2019)\citenamefont
  {Burnett}, \citenamefont {Bengtsson}, \citenamefont {Scigliuzzo},
  \citenamefont {Niepce}, \citenamefont {Kudra}, \citenamefont {Delsing},\ and\
  \citenamefont {Bylander}}]{Burnett2019}%
  \BibitemOpen
  \bibfield  {author} {\bibinfo {author} {\bibfnamefont {J.~J.}\ \bibnamefont
  {Burnett}}, \bibinfo {author} {\bibfnamefont {A.}~\bibnamefont {Bengtsson}},
  \bibinfo {author} {\bibfnamefont {M.}~\bibnamefont {Scigliuzzo}}, \bibinfo
  {author} {\bibfnamefont {D.}~\bibnamefont {Niepce}}, \bibinfo {author}
  {\bibfnamefont {M.}~\bibnamefont {Kudra}}, \bibinfo {author} {\bibfnamefont
  {P.}~\bibnamefont {Delsing}},\ and\ \bibinfo {author} {\bibfnamefont
  {J.}~\bibnamefont {Bylander}},\ }\bibfield  {title} {\bibinfo {title}
  {{Decoherence benchmarking of superconducting qubits}},\ }\href
  {https://doi.org/10.1038/s41534-019-0168-5} {\bibfield  {journal} {\bibinfo
  {journal} {npj Quantum Information}\ }\textbf {\bibinfo {volume} {5}},\
  \bibinfo {pages} {54} (\bibinfo {year} {2019})}\BibitemShut {NoStop}%
\bibitem [{Note1()}]{Note1}%
  \BibitemOpen
  \bibinfo {note} {The source code developed for this work can be found online
  at \protect \url {https://gitlab.com/DQMLab/TLSInfer.jl}.}\BibitemShut
  {Stop}%
\bibitem [{Note2()}]{Note2}%
  \BibitemOpen
  \bibinfo {note} {In the experiment we use for~$\protect \tilde {P_e}$ a
  function that takes into account relaxation as well as measurement
  visibility. For the relaxation, we use Eq.~(17) in Ref.~\protect \rev@citealp
  {Stenberg2014}. To account for measurement visibility, we clamp the
  theoretical probability between~$0.05$ and $0.95$.}\BibitemShut {Stop}%
\bibitem [{\citenamefont {Murray}\ \emph {et~al.}(2016)\citenamefont {Murray},
  \citenamefont {Lee},\ and\ \citenamefont {Jacob}}]{Murray2016}%
  \BibitemOpen
  \bibfield  {author} {\bibinfo {author} {\bibfnamefont {L.~M.}\ \bibnamefont
  {Murray}}, \bibinfo {author} {\bibfnamefont {A.}~\bibnamefont {Lee}},\ and\
  \bibinfo {author} {\bibfnamefont {P.~E.}\ \bibnamefont {Jacob}},\ }\bibfield
  {title} {\bibinfo {title} {{Parallel Resampling in the Particle Filter}},\
  }\href {https://doi.org/10.1080/10618600.2015.1062015} {\bibfield  {journal}
  {\bibinfo  {journal} {Journal of Computational and Graphical Statistics}\
  }\textbf {\bibinfo {volume} {25}},\ \bibinfo {pages} {789} (\bibinfo {year}
  {2016})}\BibitemShut {NoStop}%
\bibitem [{\citenamefont {Bertet}\ \emph {et~al.}(2006)\citenamefont {Bertet},
  \citenamefont {Harmans},\ and\ \citenamefont {Mooij}}]{Bertet2006}%
  \BibitemOpen
  \bibfield  {author} {\bibinfo {author} {\bibfnamefont {P.}~\bibnamefont
  {Bertet}}, \bibinfo {author} {\bibfnamefont {C.~J. P.~M.}\ \bibnamefont
  {Harmans}},\ and\ \bibinfo {author} {\bibfnamefont {J.~E.}\ \bibnamefont
  {Mooij}},\ }\bibfield  {title} {\bibinfo {title} {{Parametric coupling for
  superconducting qubits}},\ }\href
  {https://doi.org/10.1103/PhysRevB.73.064512} {\bibfield  {journal} {\bibinfo
  {journal} {Physical Review B}\ }\textbf {\bibinfo {volume} {73}},\ \bibinfo
  {pages} {064512} (\bibinfo {year} {2006})}\BibitemShut {NoStop}%
\bibitem [{\citenamefont {McKay}\ \emph {et~al.}(2016)\citenamefont {McKay},
  \citenamefont {Filipp}, \citenamefont {Mezzacapo}, \citenamefont {Magesan},
  \citenamefont {Chow},\ and\ \citenamefont {Gambetta}}]{McKay2016}%
  \BibitemOpen
  \bibfield  {author} {\bibinfo {author} {\bibfnamefont {D.~C.}\ \bibnamefont
  {McKay}}, \bibinfo {author} {\bibfnamefont {S.}~\bibnamefont {Filipp}},
  \bibinfo {author} {\bibfnamefont {A.}~\bibnamefont {Mezzacapo}}, \bibinfo
  {author} {\bibfnamefont {E.}~\bibnamefont {Magesan}}, \bibinfo {author}
  {\bibfnamefont {J.~M.}\ \bibnamefont {Chow}},\ and\ \bibinfo {author}
  {\bibfnamefont {J.~M.}\ \bibnamefont {Gambetta}},\ }\bibfield  {title}
  {\bibinfo {title} {{Universal Gate for Fixed-Frequency Qubits via a Tunable
  Bus}},\ }\href {https://doi.org/10.1103/PhysRevApplied.6.064007} {\bibfield
  {journal} {\bibinfo  {journal} {Physical Review Applied}\ }\textbf {\bibinfo
  {volume} {6}},\ \bibinfo {pages} {064007} (\bibinfo {year}
  {2016})}\BibitemShut {NoStop}%
\bibitem [{\citenamefont {Caldwell}\ \emph {et~al.}(2018)\citenamefont
  {Caldwell}, \citenamefont {Didier}, \citenamefont {Ryan}, \citenamefont
  {Sete}, \citenamefont {Hudson}, \citenamefont {Karalekas}, \citenamefont
  {Manenti}, \citenamefont {da~Silva}, \citenamefont {Sinclair}, \citenamefont
  {Acala}, \citenamefont {Alidoust}, \citenamefont {Angeles}, \citenamefont
  {Bestwick}, \citenamefont {Block}, \citenamefont {Bloom}, \citenamefont
  {Bradley}, \citenamefont {Bui}, \citenamefont {Capelluto}, \citenamefont
  {Chilcott}, \citenamefont {Cordova}, \citenamefont {Crossman}, \citenamefont
  {Curtis}, \citenamefont {Deshpande}, \citenamefont {Bouayadi}, \citenamefont
  {Girshovich}, \citenamefont {Hong}, \citenamefont {Kuang}, \citenamefont
  {Lenihan}, \citenamefont {Manning}, \citenamefont {Marchenkov}, \citenamefont
  {Marshall}, \citenamefont {Maydra}, \citenamefont {Mohan}, \citenamefont
  {O'Brien}, \citenamefont {Osborn}, \citenamefont {Otterbach}, \citenamefont
  {Papageorge}, \citenamefont {Paquette}, \citenamefont {Pelstring},
  \citenamefont {Polloreno}, \citenamefont {Prawiroatmodjo}, \citenamefont
  {Rawat}, \citenamefont {Reagor}, \citenamefont {Renzas}, \citenamefont
  {Rubin}, \citenamefont {Russell}, \citenamefont {Rust}, \citenamefont
  {Scarabelli}, \citenamefont {Scheer}, \citenamefont {Selvanayagam},
  \citenamefont {Smith}, \citenamefont {Staley}, \citenamefont {Suska},
  \citenamefont {Tezak}, \citenamefont {Thompson}, \citenamefont {To},
  \citenamefont {Vahidpour}, \citenamefont {Vodrahalli}, \citenamefont
  {Whyland}, \citenamefont {Yadav}, \citenamefont {Zeng},\ and\ \citenamefont
  {Rigetti}}]{Caldwell2018}%
  \BibitemOpen
  \bibfield  {author} {\bibinfo {author} {\bibfnamefont {S.~A.}\ \bibnamefont
  {Caldwell}}, \bibinfo {author} {\bibfnamefont {N.}~\bibnamefont {Didier}},
  \bibinfo {author} {\bibfnamefont {C.~A.}\ \bibnamefont {Ryan}}, \bibinfo
  {author} {\bibfnamefont {E.~A.}\ \bibnamefont {Sete}}, \bibinfo {author}
  {\bibfnamefont {A.}~\bibnamefont {Hudson}}, \bibinfo {author} {\bibfnamefont
  {P.}~\bibnamefont {Karalekas}}, \bibinfo {author} {\bibfnamefont
  {R.}~\bibnamefont {Manenti}}, \bibinfo {author} {\bibfnamefont {M.~P.}\
  \bibnamefont {da~Silva}}, \bibinfo {author} {\bibfnamefont {R.}~\bibnamefont
  {Sinclair}}, \bibinfo {author} {\bibfnamefont {E.}~\bibnamefont {Acala}},
  \bibinfo {author} {\bibfnamefont {N.}~\bibnamefont {Alidoust}}, \bibinfo
  {author} {\bibfnamefont {J.}~\bibnamefont {Angeles}}, \bibinfo {author}
  {\bibfnamefont {A.}~\bibnamefont {Bestwick}}, \bibinfo {author}
  {\bibfnamefont {M.}~\bibnamefont {Block}}, \bibinfo {author} {\bibfnamefont
  {B.}~\bibnamefont {Bloom}}, \bibinfo {author} {\bibfnamefont
  {A.}~\bibnamefont {Bradley}}, \bibinfo {author} {\bibfnamefont
  {C.}~\bibnamefont {Bui}}, \bibinfo {author} {\bibfnamefont {L.}~\bibnamefont
  {Capelluto}}, \bibinfo {author} {\bibfnamefont {R.}~\bibnamefont {Chilcott}},
  \bibinfo {author} {\bibfnamefont {J.}~\bibnamefont {Cordova}}, \bibinfo
  {author} {\bibfnamefont {G.}~\bibnamefont {Crossman}}, \bibinfo {author}
  {\bibfnamefont {M.}~\bibnamefont {Curtis}}, \bibinfo {author} {\bibfnamefont
  {S.}~\bibnamefont {Deshpande}}, \bibinfo {author} {\bibfnamefont {T.~E.}\
  \bibnamefont {Bouayadi}}, \bibinfo {author} {\bibfnamefont {D.}~\bibnamefont
  {Girshovich}}, \bibinfo {author} {\bibfnamefont {S.}~\bibnamefont {Hong}},
  \bibinfo {author} {\bibfnamefont {K.}~\bibnamefont {Kuang}}, \bibinfo
  {author} {\bibfnamefont {M.}~\bibnamefont {Lenihan}}, \bibinfo {author}
  {\bibfnamefont {T.}~\bibnamefont {Manning}}, \bibinfo {author} {\bibfnamefont
  {A.}~\bibnamefont {Marchenkov}}, \bibinfo {author} {\bibfnamefont
  {J.}~\bibnamefont {Marshall}}, \bibinfo {author} {\bibfnamefont
  {R.}~\bibnamefont {Maydra}}, \bibinfo {author} {\bibfnamefont
  {Y.}~\bibnamefont {Mohan}}, \bibinfo {author} {\bibfnamefont
  {W.}~\bibnamefont {O'Brien}}, \bibinfo {author} {\bibfnamefont
  {C.}~\bibnamefont {Osborn}}, \bibinfo {author} {\bibfnamefont
  {J.}~\bibnamefont {Otterbach}}, \bibinfo {author} {\bibfnamefont
  {A.}~\bibnamefont {Papageorge}}, \bibinfo {author} {\bibfnamefont {J.-P.}\
  \bibnamefont {Paquette}}, \bibinfo {author} {\bibfnamefont {M.}~\bibnamefont
  {Pelstring}}, \bibinfo {author} {\bibfnamefont {A.}~\bibnamefont
  {Polloreno}}, \bibinfo {author} {\bibfnamefont {G.}~\bibnamefont
  {Prawiroatmodjo}}, \bibinfo {author} {\bibfnamefont {V.}~\bibnamefont
  {Rawat}}, \bibinfo {author} {\bibfnamefont {M.}~\bibnamefont {Reagor}},
  \bibinfo {author} {\bibfnamefont {R.}~\bibnamefont {Renzas}}, \bibinfo
  {author} {\bibfnamefont {N.}~\bibnamefont {Rubin}}, \bibinfo {author}
  {\bibfnamefont {D.}~\bibnamefont {Russell}}, \bibinfo {author} {\bibfnamefont
  {M.}~\bibnamefont {Rust}}, \bibinfo {author} {\bibfnamefont {D.}~\bibnamefont
  {Scarabelli}}, \bibinfo {author} {\bibfnamefont {M.}~\bibnamefont {Scheer}},
  \bibinfo {author} {\bibfnamefont {M.}~\bibnamefont {Selvanayagam}}, \bibinfo
  {author} {\bibfnamefont {R.}~\bibnamefont {Smith}}, \bibinfo {author}
  {\bibfnamefont {A.}~\bibnamefont {Staley}}, \bibinfo {author} {\bibfnamefont
  {M.}~\bibnamefont {Suska}}, \bibinfo {author} {\bibfnamefont
  {N.}~\bibnamefont {Tezak}}, \bibinfo {author} {\bibfnamefont {D.~C.}\
  \bibnamefont {Thompson}}, \bibinfo {author} {\bibfnamefont {T.-W.}\
  \bibnamefont {To}}, \bibinfo {author} {\bibfnamefont {M.}~\bibnamefont
  {Vahidpour}}, \bibinfo {author} {\bibfnamefont {N.}~\bibnamefont
  {Vodrahalli}}, \bibinfo {author} {\bibfnamefont {T.}~\bibnamefont {Whyland}},
  \bibinfo {author} {\bibfnamefont {K.}~\bibnamefont {Yadav}}, \bibinfo
  {author} {\bibfnamefont {W.}~\bibnamefont {Zeng}},\ and\ \bibinfo {author}
  {\bibfnamefont {C.}~\bibnamefont {Rigetti}},\ }\bibfield  {title} {\bibinfo
  {title} {{Parametrically Activated Entangling Gates Using Transmon Qubits}},\
  }\href {https://doi.org/10.1103/PhysRevApplied.10.034050} {\bibfield
  {journal} {\bibinfo  {journal} {Physical Review Applied}\ }\textbf {\bibinfo
  {volume} {10}},\ \bibinfo {pages} {034050} (\bibinfo {year}
  {2018})}\BibitemShut {NoStop}%
\bibitem [{\citenamefont {Klimov}\ \emph {et~al.}(2020)\citenamefont {Klimov},
  \citenamefont {Kelly}, \citenamefont {Martinis},\ and\ \citenamefont
  {Neven}}]{Klimov2020}%
  \BibitemOpen
  \bibfield  {author} {\bibinfo {author} {\bibfnamefont {P.~V.}\ \bibnamefont
  {Klimov}}, \bibinfo {author} {\bibfnamefont {J.}~\bibnamefont {Kelly}},
  \bibinfo {author} {\bibfnamefont {J.~M.}\ \bibnamefont {Martinis}},\ and\
  \bibinfo {author} {\bibfnamefont {H.}~\bibnamefont {Neven}},\ }\bibfield
  {title} {\bibinfo {title} {{The Snake Optimizer for Learning Quantum
  Processor Control Parameters}},\ }\Eprint {https://arxiv.org/abs/2006.04594}
  {arXiv:2006.04594}  (\bibinfo {year} {2020})\BibitemShut {NoStop}%
\bibitem [{\citenamefont {B{\'{e}}janin}\ \emph {et~al.}(2016)\citenamefont
  {B{\'{e}}janin}, \citenamefont {McConkey}, \citenamefont {Rinehart},
  \citenamefont {Earnest}, \citenamefont {McRae}, \citenamefont {Shiri},
  \citenamefont {Bateman}, \citenamefont {Rohanizadegan}, \citenamefont
  {Penava}, \citenamefont {Breul}, \citenamefont {Royak}, \citenamefont
  {Zapatka}, \citenamefont {Fowler},\ and\ \citenamefont
  {Mariantoni}}]{Bejanin2016}%
  \BibitemOpen
  \bibfield  {author} {\bibinfo {author} {\bibfnamefont {J.~H.}\ \bibnamefont
  {B{\'{e}}janin}}, \bibinfo {author} {\bibfnamefont {T.~G.}\ \bibnamefont
  {McConkey}}, \bibinfo {author} {\bibfnamefont {J.~R.}\ \bibnamefont
  {Rinehart}}, \bibinfo {author} {\bibfnamefont {C.~T.}\ \bibnamefont
  {Earnest}}, \bibinfo {author} {\bibfnamefont {C.~R.~H.}\ \bibnamefont
  {McRae}}, \bibinfo {author} {\bibfnamefont {D.}~\bibnamefont {Shiri}},
  \bibinfo {author} {\bibfnamefont {J.~D.}\ \bibnamefont {Bateman}}, \bibinfo
  {author} {\bibfnamefont {Y.}~\bibnamefont {Rohanizadegan}}, \bibinfo {author}
  {\bibfnamefont {B.}~\bibnamefont {Penava}}, \bibinfo {author} {\bibfnamefont
  {P.}~\bibnamefont {Breul}}, \bibinfo {author} {\bibfnamefont
  {S.}~\bibnamefont {Royak}}, \bibinfo {author} {\bibfnamefont
  {M.}~\bibnamefont {Zapatka}}, \bibinfo {author} {\bibfnamefont {A.~G.}\
  \bibnamefont {Fowler}},\ and\ \bibinfo {author} {\bibfnamefont
  {M.}~\bibnamefont {Mariantoni}},\ }\bibfield  {title} {\bibinfo {title}
  {{Three-Dimensional Wiring for Extensible Quantum Computing: The Quantum
  Socket}},\ }\href {https://doi.org/10.1103/PhysRevApplied.6.044010}
  {\bibfield  {journal} {\bibinfo  {journal} {Physical Review Applied}\
  }\textbf {\bibinfo {volume} {6}},\ \bibinfo {pages} {044010} (\bibinfo {year}
  {2016})}\BibitemShut {NoStop}%
\bibitem [{\citenamefont {Moghaddam}\ \emph {et~al.}(2019)\citenamefont
  {Moghaddam}, \citenamefont {Chang}, \citenamefont {Nsanzineza}, \citenamefont
  {Vadiraj},\ and\ \citenamefont {Wilson}}]{Moghaddam2019}%
  \BibitemOpen
  \bibfield  {author} {\bibinfo {author} {\bibfnamefont {M.~V.}\ \bibnamefont
  {Moghaddam}}, \bibinfo {author} {\bibfnamefont {C.~W.~S.}\ \bibnamefont
  {Chang}}, \bibinfo {author} {\bibfnamefont {I.}~\bibnamefont {Nsanzineza}},
  \bibinfo {author} {\bibfnamefont {A.~M.}\ \bibnamefont {Vadiraj}},\ and\
  \bibinfo {author} {\bibfnamefont {C.~M.}\ \bibnamefont {Wilson}},\ }\bibfield
   {title} {\bibinfo {title} {{Carbon nanotube-based lossy transmission line
  filter for superconducting qubit measurements}},\ }\href
  {https://doi.org/10.1063/1.5116109} {\bibfield  {journal} {\bibinfo
  {journal} {Applied Physics Letters}\ }\textbf {\bibinfo {volume} {115}},\
  \bibinfo {pages} {213504} (\bibinfo {year} {2019})}\BibitemShut {NoStop}%
\bibitem [{\citenamefont {Santavicca}\ and\ \citenamefont
  {Prober}(2008)}]{Santavicca2008}%
  \BibitemOpen
  \bibfield  {author} {\bibinfo {author} {\bibfnamefont {D.~F.}\ \bibnamefont
  {Santavicca}}\ and\ \bibinfo {author} {\bibfnamefont {D.~E.}\ \bibnamefont
  {Prober}},\ }\bibfield  {title} {\bibinfo {title} {{Impedance-matched
  low-pass stripline filters}},\ }\href
  {https://doi.org/10.1088/0957-0233/19/8/087001} {\bibfield  {journal}
  {\bibinfo  {journal} {Measurement Science and Technology}\ }\textbf {\bibinfo
  {volume} {19}},\ \bibinfo {pages} {087001} (\bibinfo {year}
  {2008})}\BibitemShut {NoStop}%
\end{thebibliography}
\end{document}